\documentclass[3p]{elsarticle}
\pdfoutput=1

\usepackage{url}
\usepackage{indentfirst}
\usepackage{gensymb}
\usepackage{graphicx}
\usepackage{pdfpages}
\usepackage{wrapfig}
\usepackage{float}
\usepackage[caption = false]{subfig}
\usepackage{amsmath,bm}
\usepackage{titlesec}
\usepackage{fancyhdr}
\usepackage{color, colortbl}
\usepackage{graphicx}
\usepackage{multirow}
\usepackage{array}
\usepackage{ragged2e}
\usepackage[toc,page]{appendix}

\definecolor{Gray}{gray}{0.9}
\definecolor{White}{gray}{1}

\newcolumntype{?}{!{\vrule width 1pt}}

\graphicspath{{Figures/}}

\begin{document}

	\begin{frontmatter}
		
		\title{Grain-Resolved Temperature-Dependent Anisotropy in Hexagonal Ti-7Al Revealed by Synchrotron X-Ray Dif{}fraction}
		
		
		\author[1]{Rachel Lim\corref{mycorrespondingauthor}}
		\ead{relim@andrew.cmu.edu} 
		
		\author[2]{Darren C. Pagan}
		\author[3]{Donald E. Boyce}
		\author[4]{Joel V. Bernier}
		\author[5]{Paul A. Shade}
		\author[1]{Anthony D. Rollett}

		\cortext[mycorrespondingauthor]{Corresponding author}

		\address[1]{Department of Materials Science and Engineering, Carnegie Mellon University, Pittsburgh, PA 15213, USA}
		\address[2]{Cornell High Energy Synchrotron Source, Ithaca, NY 14853, USA}
		\address[3]{Cornell University, Ithaca, NY 14850, USA}
		\address[4]{Lawrence Livermore National Laboratory, Livermore, CA 94550, USA}
		\address[5]{Air Force Research Laboratory, Wright-Patterson AFB, OH 45433, USA}
		
		\begin{abstract}
			Hexagonal metals have anisotropic coef{}ficients of thermal expansion causing grain-level internal stresses during heating. High energy x-ray dif{}fraction microscopy, a non-destructive, in situ, micromechanical and microstructural characterization technique, has been used to determine the anisotropic coef{}ficients of thermal expansion (CTEs) for Ti-7Al. Two samples of polycrystalline $\alpha$-phase Ti-7Al were continuously heated from room temperature to 850~$\degree$C while far-field HEDM scans were collected. The results showed a change in the ratio of the CTEs in the `a' and `c' directions which explains discrepancies found in the literature. The CTE additionally appears to be af{}fected by the dissolution of $\alpha_2$ precipitates. Analysis of the grain-resolved micromechanical data also shows reconfiguration of the grain scale stresses likely due to anisotropic expansion driving crystallographic slip.
		\end{abstract}
		
		
	\end{frontmatter}

\section{Introduction}
	The ef{}fects of thermal anisotropy were noted in 1944 by Boas and Honeycombe \cite{Boas1944} who stated that ``It appears, therefore, that metals which possess a high degree of anisotropy of thermal expansion cannot be obtained in a stress-free condition at room temperature by casting or annealing." Thus, in hexagonal metals, anisotropic coef{}ficients of thermal expansion (CTEs) are believed to generate the majority of type II residual stresses, i.e., microscale stresses which remain even after the removal of all external loads \cite{KesavanNair1995}, during cooling from high-temperature processing conditions \cite{Warwick2012,Zheng2019}.  These type II residual stresses af{}fect the local variations in stress, and therefore can play a critical role in either accelerating or suppressing the development of damage that leads to premature failure. Thus, the prediction of failure initiation hinges on residual stress instantiation that in turn relies on accurate CTE values (and their ratios). However, there is little agreement in the literature on the CTEs for these metals \cite{Pawar1968}, and notably the reported CTEs for titanium (Table~\ref{tab:Ti_CTE}) vary from 9.26~$\times$~10\textsuperscript{-6}/$\degree$C to 13.17~$\times$~10\textsuperscript{-6}/$\degree$C, lacking agreement even on the ratio between the CTEs in the `a' and `c' directions ($\alpha_a$ and $\alpha_c$), all of which motivated a more detailed examination of this phenomenon.

\begin{table}[b!]
	\centering
	\begin{tabular}{c|c|c|c|c|c}
		\textbf{Year} & \textbf{Paper} & \textbf{$\alpha_a$}  & \textbf{$\alpha_c$} & \textbf{$\alpha_v$} & \textbf{Temp. Range ($\degree$C)} \\ \hline
		
		\hline
		1942 & Erfling \cite{Erfling1942} &  &  & 8.24 & 20-40 \\ 
		\hline
		1949 & Greiner \& Ellis \cite{Greiner1949} &   &  & 9 & 30-200 \\ 
		\hline
		1953 & McHargue \& Hammond \cite{Mchargue1953}& 11 & 8.8 & 10.3 & 25-225\\ 
		\hline
		1953 & Berry \& Raynor \cite{Berry1953}& 11.03 & 13.37 & 11.81 & r.t.-700\\ 
		\hline
		1959 & Spreadborough \& Christian \cite{Spreadborough1959} & 9.55 & 10.65 & 9.92 & 0-600 \\ 
		\hline
		1962 & Roberts \cite{Roberts1962} & 10 & 9.95 &  & \\ 
		\hline
		1962 & Willens \cite{Willens1962}& 9.41 & 11.18 &  & 0-400 \\ 
		\hline
		1968 & Pawar \& Deshpande \cite{Pawar1968}& 9.5 & 5.6 & 8.2 & 28-155\\ 
		\hline
		1975 & Touloukian \cite{Touloukian1975}& 9.23 & 9.57 &  & \\ 
		\hline
		2019 & Zheng et al.* \cite{Zheng2019}& ** & *** &  & r.t.-850\\ 
	\end{tabular}
	\justifying
	
	\noindent\** Extracted from a simulation \par
	\noindent** $5.23\times10^{-19}~T^5 - 7.60\times10^{-16}~T^4 + 3.84\times10^{-13}~T^3 - 5.47\times10^{-11}~T^2 +8.39\times10^{-9}~T + 1.91\times10^{-6}$ \par
	\noindent*** $7.76\times10^{-17}~T^4 - 8.95\times10^{-14}~T^3 + 5.46\times10^{-11}~T^2 + 2.46\times10^{-10}~T + 3.00\times10^{-6} $
	\caption{CTEs reported in literature (all CTE values except Zheng et al. are $\times$10\textsuperscript{-6}/$\degree$C).}
	\label{tab:Ti_CTE}
\end{table}

	Historically, thermal expansion of polycrystals has been studied using dilatometry \cite{Roberts1962, ASMInternational2002} where a rod-shaped specimen is placed in a furnace, and the change in length is measured as a function of temperature. However, this technique is not capable of probing the expansion along specific crystal directions, and only the ef{}fective macroscale thermal properties of the aggregate can be extracted. In order to study the anisotropic linear thermal expansion of a material, the characterization technique used must be capable of measuring expansion along dif{}ferent crystallographic directions. Powder dif{}fraction \cite{Spreadborough1959, Pawar1968} is commonly employed to obtain the variation in lattice parameter(s), which is then used to calculate the CTEs. The dif{}fraction measurements provide information on lattice plane spacings, from which lattice strains, due to both thermal and mechanical strains, can be calculated.
	
	Anisotropic lattice expansion at the microscale becomes coupled to mechanical response as the material attempts to maintain local compatibility and stress equilibrium. Thermal expansion is equivalent to eigenstrain, i.e., stress-free strain, and for titanium and other hexagonal metals, the dif{}ference between the a-axis and c-axis expansion coupled with the variations in local crystallographic orientation results in the generation of elastic strains (and stresses) \cite{Boab1944}. These mechanical strains are particularly dif{}ficult to decouple from thermal strains using powder dif{}fraction methods. Thus, we use far-field high energy x-ray dif{}fraction microscopy (f{}f-HEDM) measurements with \textit{in situ} heating to extract the strain tensor for individual crystals within a polycrystalline aggregate as a function of temperature.
	
	The f{}f-HEDM technique measures total lattice strain and cannot distinguish between the thermal and mechanical contributions, although the ability to measure full lattice strain tensors allows some decoupling of thermal and mechanical strains to be performed. We assume that non-zero shear strains and dispersion observed in normal strain components between grains is most directly the result of the development of mechanical strains, while the average strain across the ensemble of grains is the thermal strain. It should be noted that these variations in strains between individual grains due to neighborhood constraints cannot be isolated with powder dif{}fraction but is a direct result from f{}f-HEDM. This in turn allows for better understanding of the micromechanical interactions and subsequent interpretation of CTE measurements in a polycrystal.
	
	The Ti-7Al studied in this work is a hexagonal close packed $\alpha$-Ti alloy, and is similar to the $\alpha$-phase in Ti-6Al-4V, which is a commonly used alloy in the aerospace and biomedical industries. Ti-7Al  is thermally anisotropic, making it interesting for thermal expansion and residual stress studies \cite{Spreadborough1959, Pawar1968, Russell1997}, and a significant amount of previous work has been done on characterizing the deformation of Ti-7Al using HEDM \cite{Lienert2009,Schuren2015,Chatterjee2016,Turner2017,Pagan2017,Pagan2018,Nygren2019,Nygren2020}. Plastic deformation in $\alpha$-Ti mainly occurs through the slip on the basal systems $\langle$a$\rangle$, $\langle$1$\bar{2}$10$\rangle$\{0001\}, and on the prismatic systems $\langle$a$\rangle$, $\langle$1$\bar{2}$10$\rangle$\{10$\bar{1}$0\}. Additionally, slip can also be observed on the pyramidal $\langle$a$\rangle$, $\langle$11$\bar{2}$0$\rangle$\{1$\bar{1}$01\} slip systems but with lower frequency \cite{Lutjering2007,Lienert2009,Pagan2017,Tan1998,Williams2002}. These families of slip systems have been shown to vary significantly in strength \cite{Lutjering2007,Pagan2017}. Lastly, slip on the pyramidal $\langle$c+a$\rangle$ system occurs through the glide of dislocations on the \{10$\bar{1}$1\} and \{11$\bar{2}$2\} plane families \cite{Williams1968}. Although these systems are higher strength, pyramidal $\langle$c+a$\rangle$ slip will occur in some grains as it is a necessary deformation mode to close the yield surface. As plastic deformation occurs, dislocation interactions cause hardening on these planes \cite{Pagan2017}. As a consequence of the competing hardening and softening behaviors, Ti-7Al exhibits little to no hardening on a macroscopic level \cite{Pagan2017}. 
	
	In addition, much of the strength of Ti-7Al can be attributed to aging of the material which leads to short-range ordering (SRO) and the subsequent development of coherent $\alpha$$_2$ Ti$_3$Al nanoprecipitates, which af{}fects the mechanical behavior of titanium alloys with greater than 5\% aluminum \cite{VandeWalle2002,Venkataraman2017,Gardner2020}. The $\alpha$$_2$ precipitates suppress twinning and increase slip on the basal planes \cite{Fitzner2016,Neeraj2001} and strengthen the material until the precipitates are sheared through dislocation slip \cite{Pagan2017}.
	
	In this work, far-field high energy x-ray dif{}fraction microscopy (f{}f-HEDM) measurements is employed with \textit{in situ} heating to track the evolution of the micromechanical state of individual crystals within a polycrystalline aggregate as a function of temperature in order to study the the ef{}fects of anisotropy in the thermal expansion of Ti-7Al.

\section{Methods}
	Far-field high energy x-ray dif{}fraction microscopy is a non-destructive, \textit{in situ}, materials characterization technique, which can be used to track three-dimensional micromechanical evolution as a response to external stimuli. During the use of this technique, a sample is rotated about a single axis by an angle $\omega$ as the volume of interest is illuminated by the x-ray beam. When a family of planes \{\textit{hkl}\} in a grain satisfies the dif{}fraction condition, it will dif{}fract, producing a peak in intensity on the detector. Dif{}fractograms are acquired at regular intervals integrated over $\Delta\omega$ no greater than $1.0\degree$. Data collection over a full 360$\degree$ rotation range allows for observation of $\sim$50 to 100 dif{}fraction peaks for each grain. In f{}f-HEDM measurements, the detector is placed $\sim$1 m away from the sample. In this setup, the dif{}fraction peaks line up along Debye-Scherrer rings, with the small deviations of peak locations from idealized positions enabling f{}f-HEDM to determine the grain-averaged orientation, center of mass (COM), and lattice strain tensor of the grains in the illuminated region of the sample \cite{Bernier2011,Oddershede2010, Bernier2020}. 

	In addition to the anisotropy of the CTEs for titanium, it has been shown that there is a possible temperature dependence \cite{Touloukian1975,Russell1997,Zheng2019}, and this dependence of lattice parameter on temperature can be represented by $l(T)$. For a material, the CTE, $\alpha$, for a given lattice parameter can be defined as 
		\begin{equation}
		\alpha_l = \frac{\varepsilon_l}{dT} = \frac{\frac{dl}{l}}{dT} = \frac{1}{l}\frac{dl}{dT}
		\end{equation}
	thus, the temperature dependent CTE is 
		\begin{equation} \label{eq:alphaT}
		\alpha_l(T) = \frac{1}{l(T)}\frac{dl(T)}{dT} .
		\end{equation}
	
	In order to study the micromechanical evolution of the sample and the ef{}fects of thermal expansion, the total lattice strain measured by f{}f-HEDM must be decomposed into its elastic and thermal contributions.
	\begin{equation}
	\bm{\varepsilon \,}^{to} = \bm{\varepsilon \,}^{th} + \bm{\varepsilon \,}^{el} \label{eq:total lattice strain}
	\end{equation}
	where $\bm{\varepsilon \,}^{to}$ is the total lattice strain, $\bm{\varepsilon \,}^{th}$ is the thermal strain and $\bm{\varepsilon \,}^{el}$ is the elastic strain. Since f{}f-HEDM measures strain in the sample frame, it must first be transformed into the crystal reference frame using 
	\begin{equation}
		[\bm{\varepsilon \,}^{to}]_c = \mathbf{g}^T \; [\bm{\varepsilon \,}^{to}]_s \; \mathbf{g}  \label{eq:total crys samp}
	\end{equation}
	where \textbf{g} is defined as the coordinate transformation which takes a vector from the crystal frame to the sample frame, $[\bm{\varepsilon \,}^{to}]_c$ is the total lattice strain in the crystal frame, and $[\bm{\varepsilon \,}^{to}]_s$ is the total lattice strain in the sample frame. The thermal strain can be assumed to be the average of the total lattice strains in the crystal frame 
	\begin{equation}
		[\bm{\varepsilon \,}^{th}]_c = [\bar{\bm{\varepsilon \,}}^{to}]_c . \label{eq:thermal strain}
	\end{equation}
	By combining equations \ref{eq:total lattice strain} and \ref{eq:thermal strain}, we get that 
	\begin{equation}
		[\bm{\varepsilon \,}^{el}]_c = [\bm{\varepsilon \,}^{to}]_c -  [\bar{\bm{\varepsilon \,}}^{to}]_c .
	\end{equation}
	Then, we transform the elastic strain back into the sample frame with 
	\begin{equation}
		[\bm{\varepsilon \,}^{el}]_s = \mathbf{g} \;[\bm{\varepsilon \,}^{el}]_c \; \mathbf{g}^T .
	\end{equation}

\section{Experiment}
	\subsection{Material}
\begin{figure}[h!]
			\centering
			\subfloat[Sample 1]{\includegraphics[width=0.35\textwidth]{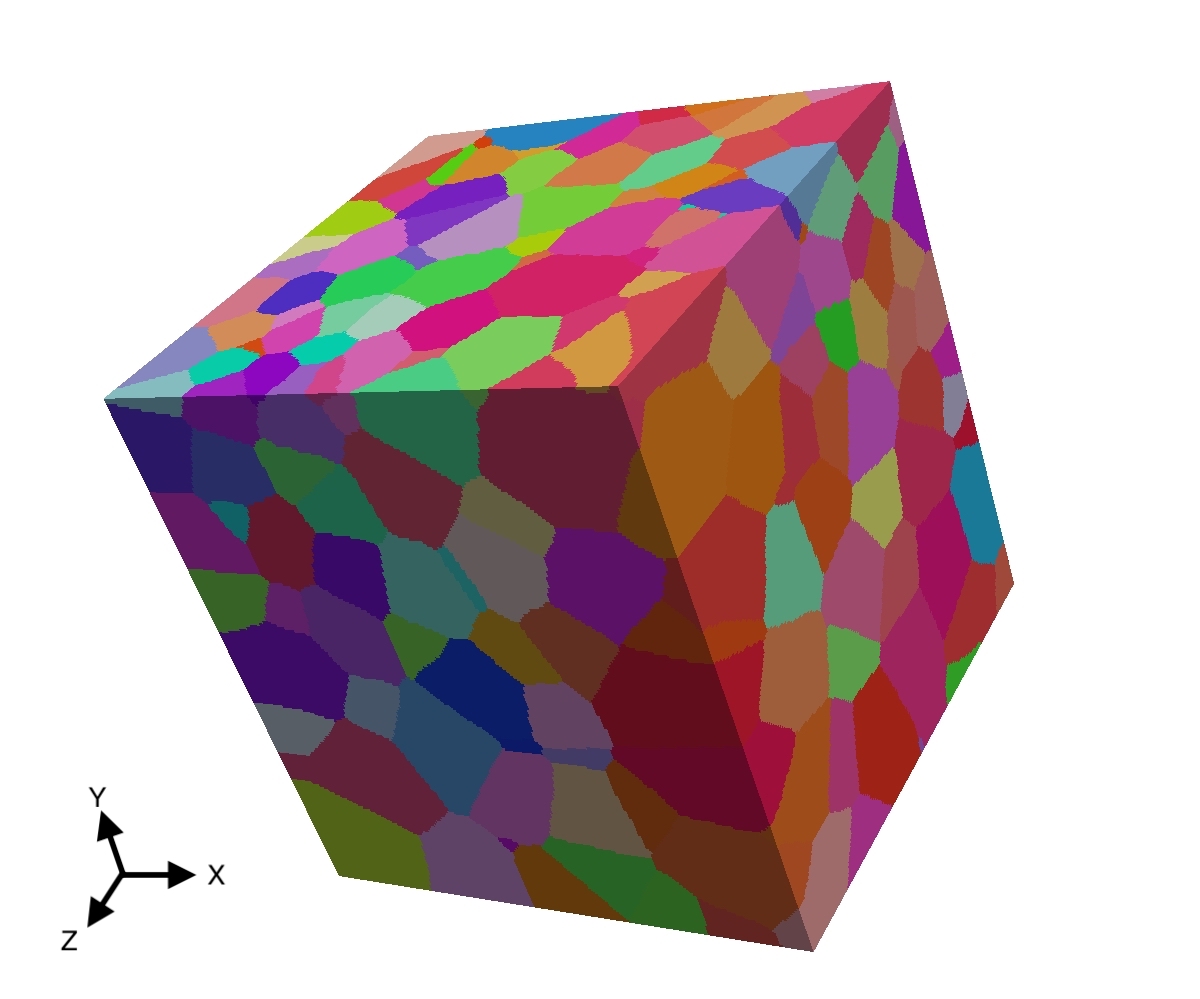}}\hspace{5pt}
			\subfloat[Sample 2]{\includegraphics[width=0.35\textwidth]{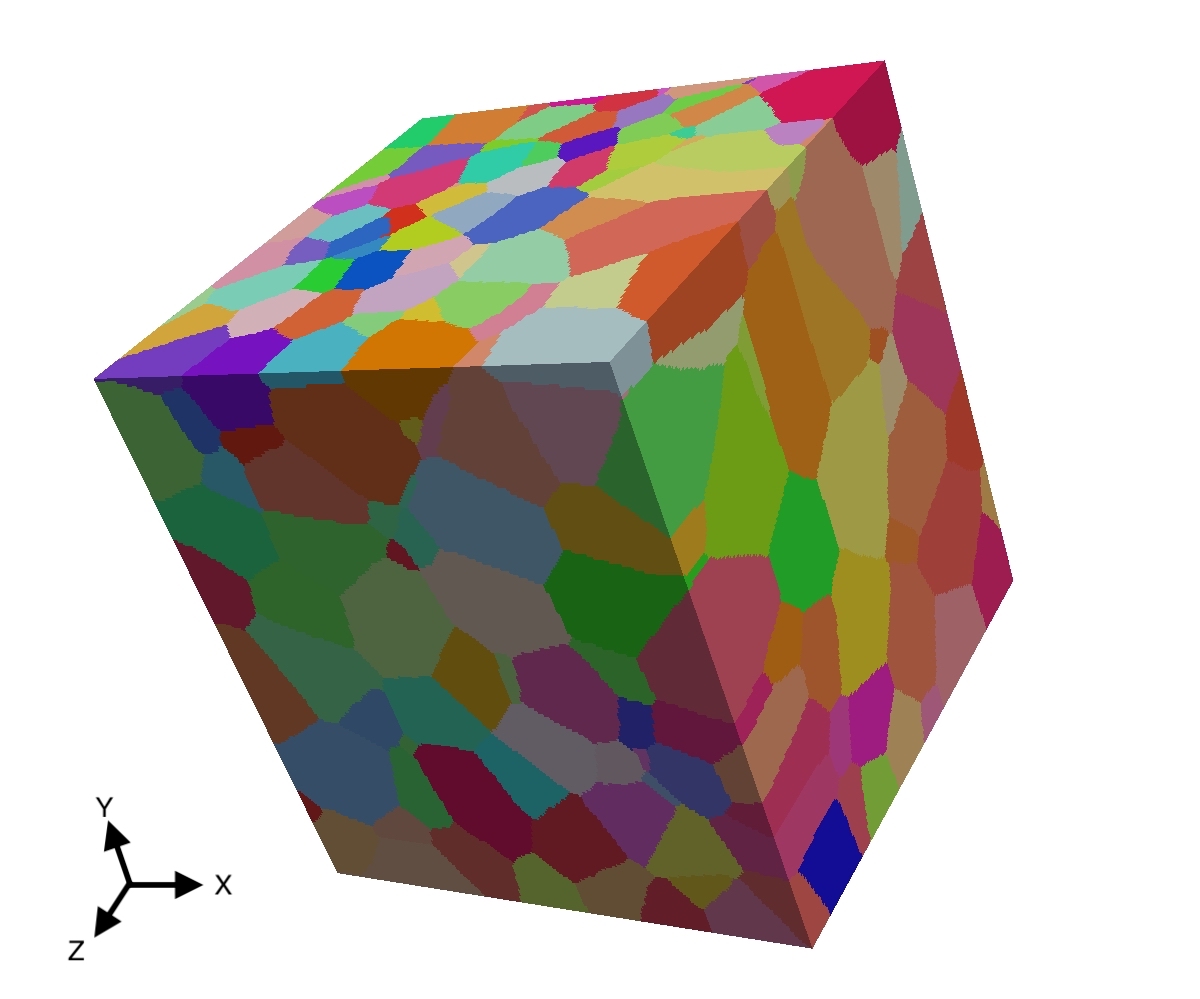}} 
			\caption{ Three-dimensional renderings of the sample volumes grown from the centroids measured in the f{}f-HEDM} 
			\label{fig:microstructures}	
		\end{figure}

		The Ti-7Al material (nominal composition Ti-7.02Al-0.11O-0.015Fe wt.\%) used for this work was cast as an ingot and hot isostatic pressed (HIP) to reduce porosity. It was extruded and then annealed at 962~$\degree$C for 24 hours before air cooling \cite{Pilchak2013}. The samples were cut using electrical discharge machining to minimize the introduction of additional residual stresses. Figures \ref{fig:microstructures}a and \ref{fig:microstructures}b show microstructural representations of the measured volumes from the two samples based on a Voronoi tessellation from the grain centers (as given by the f{}f-HEDM). The microstructures are equiaxed and have an approximate grain size of 50-100 $\mu$m where the textures of the samples were similar but not exactly the same.

	\subsection{Experiment description}
\begin{figure}[h!]
			\centering
			\includegraphics[width=0.75\textwidth]{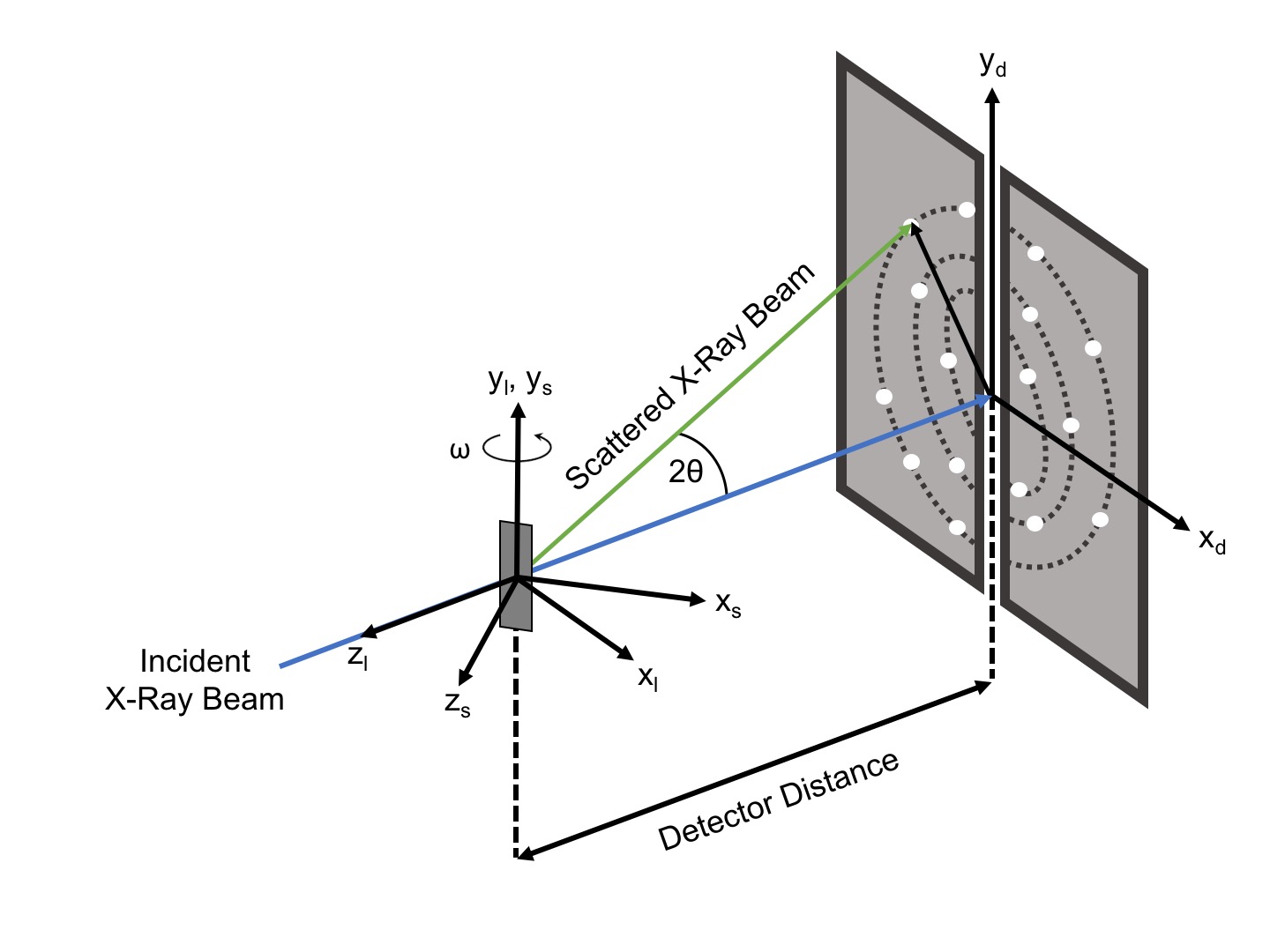}
			\caption{Schematic of the f{}f-HEDM experimental setup with two Dexela detectors side by side. The detector, laboratory, and sample frames are labeled as d, l, and s respectively, and the incident x-ray beam travels in the -z\textsubscript{d} direction.} 
			\label{fig:HEDM_setup}	
		\end{figure}

		Two thermal expansion experiments were performed on samples of Ti-7Al using the RAMS2 load frame \cite{Shade2015} at the F2 beamline at the Cornell High Energy Synchrotron Source (CHESS). Fig. \ref{fig:HEDM_setup} shows a schematic of the experimental geometry. Each sample had a gauge length of 8 mm and a 1 mm x 1 mm cross-sectional area.  The sample was heated to 850~$\degree$C at a rate of $\sim$8.5~$\degree$C/min using an x-ray transparent halogen bulb furnace with an elliptical mirror to focus the light onto the sample. The furnace was mounted onto the RAMS2 load frame as shown in Fig. 1 of Pagan et al. \cite{Pagan2018}. Far-field HEDM scans were acquired at regular intervals during heating, each with a full rotation of 360$\degree$ and an $\omega$ interval of 0.25$\degree$, using a 61.3 keV x-ray beam. A 1.1 mm tall volume (with 50 $\mu$m on the top and bottom to allow for slit scattering) of f{}f-HEDM data was collected using two Dexela 2923 detectors (3888 x 3072 pixels, 74.8 $\mu$m pixel size) mounted side-by-side (Fig. \ref{fig:HEDM_setup}).
		
	\subsection{Data processing}
		The dif{}fraction data was reduced using the HEXRD software package \url{(https://github.com/joelvbernier/hexrd)} \cite{Bernier2011}. The initial detector parameters were calibrated using powder patterns from a CeO\textsubscript{2} sample, and the grains in the Ti-7Al sample were indexed and fit using lattice parameters of $a=2.932$ {\AA} and $c=4.684$ {\AA}. Then, the detector calibration was refined using a high completeness grain close to the vertical center of the scanned volume. The reduced grain data was filtered (completeness~$>$~90\% and normalized sum of square residuals~$< 2\times10^{-3}$ \cite{Tari2018}) so that only high fidelity grains remained.

\section{Results} \label{sec:results}
\subsection{Ti-7Al coef{}ficients of thermal expansion}
	\begin{figure}[b!]
		\centering
		\subfloat{\includegraphics[width=0.4\textwidth]{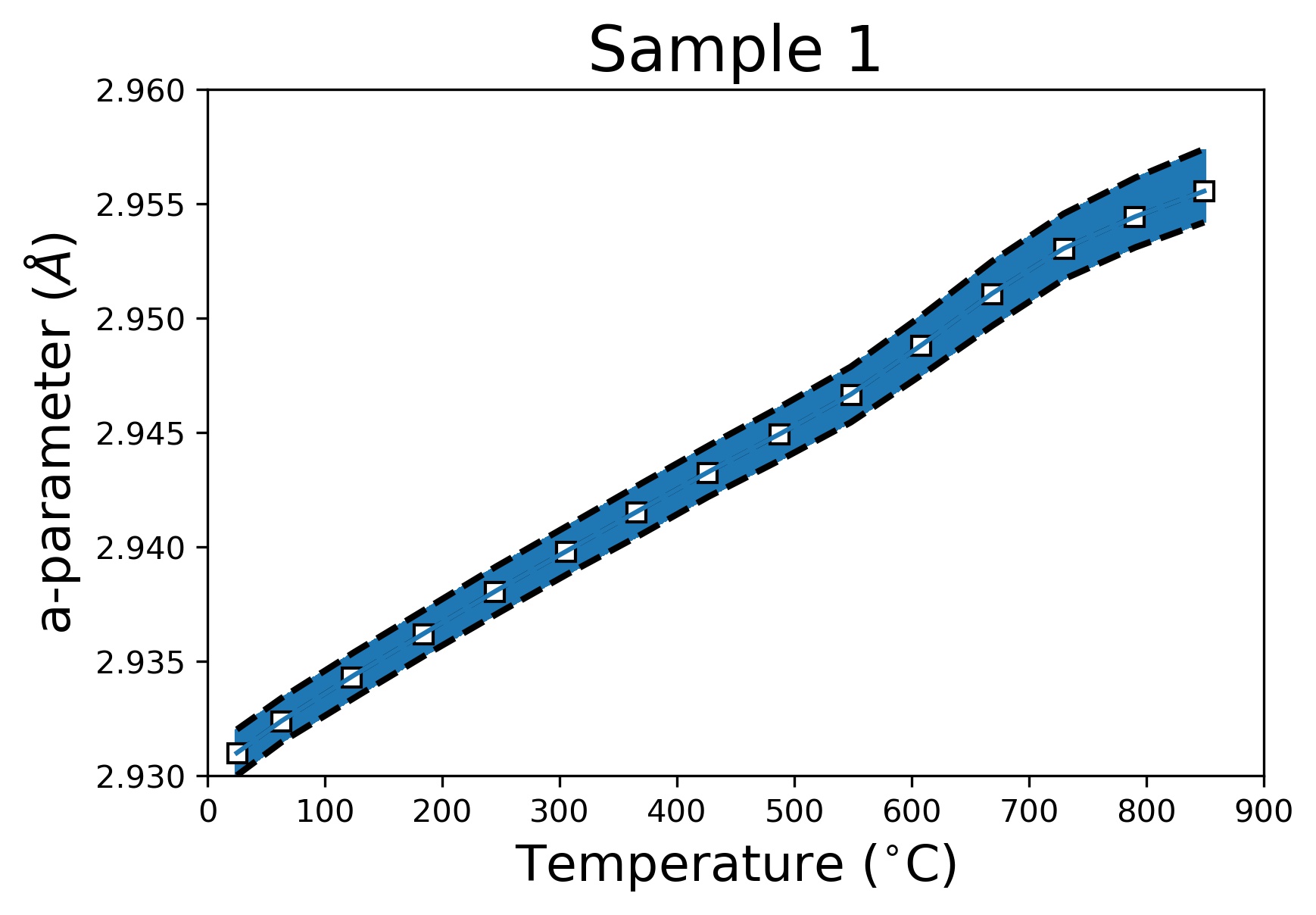}}\hspace{5pt}
		\subfloat{\includegraphics[width=0.4\textwidth]{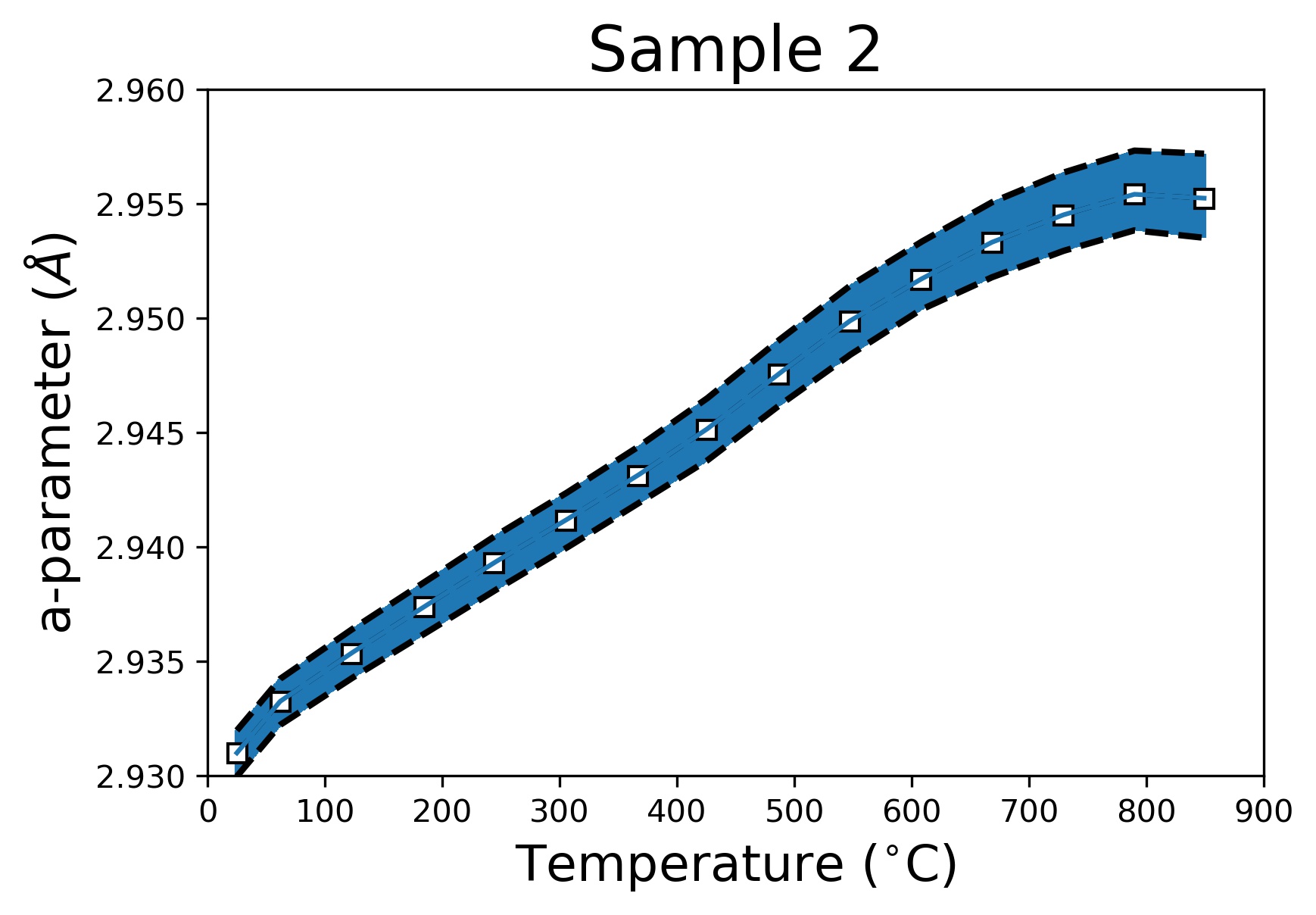}} \\
		\subfloat{\includegraphics[width=0.4\textwidth]{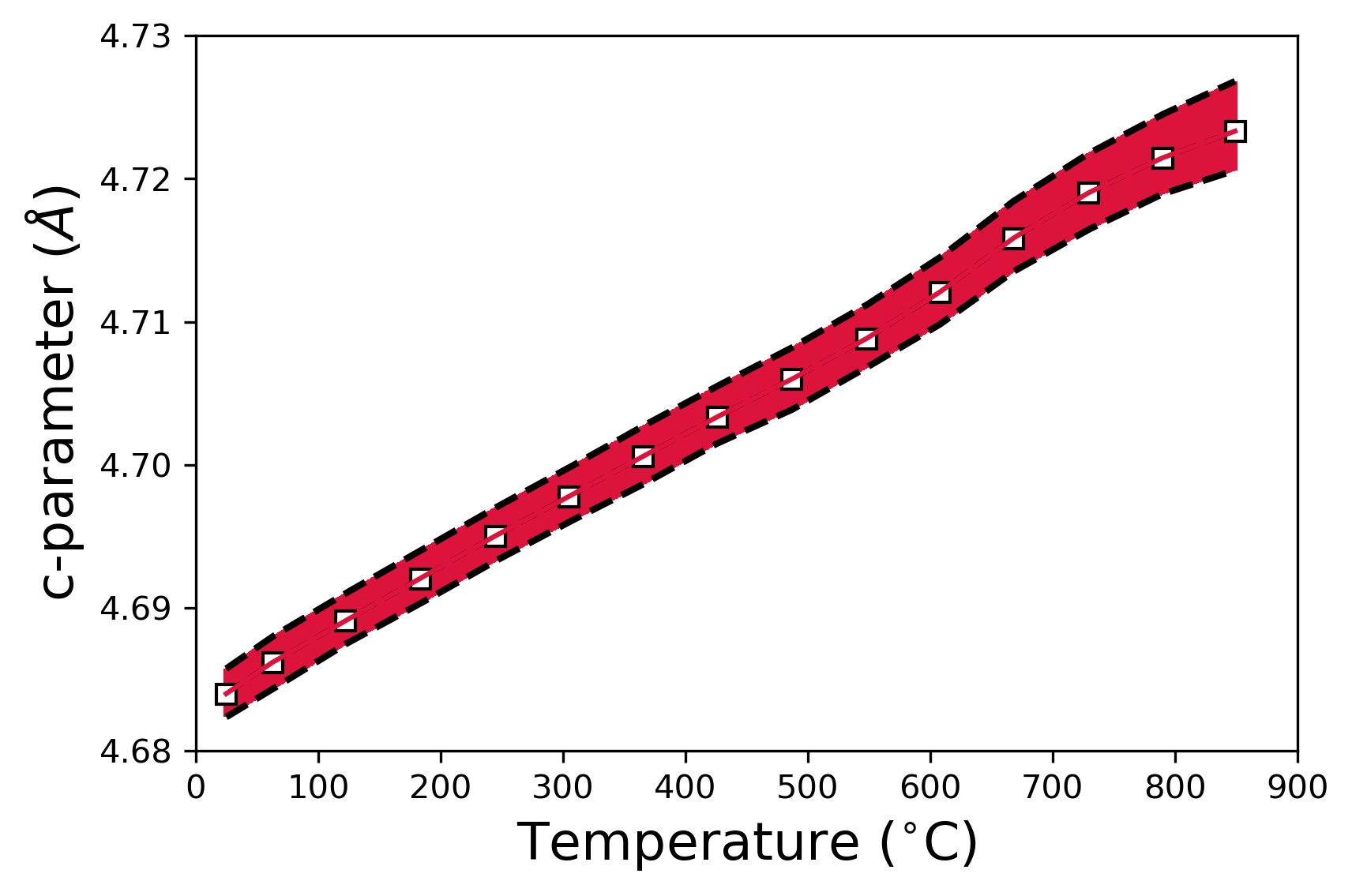}}\hspace{5pt}
		\subfloat{\includegraphics[width=0.4\textwidth]{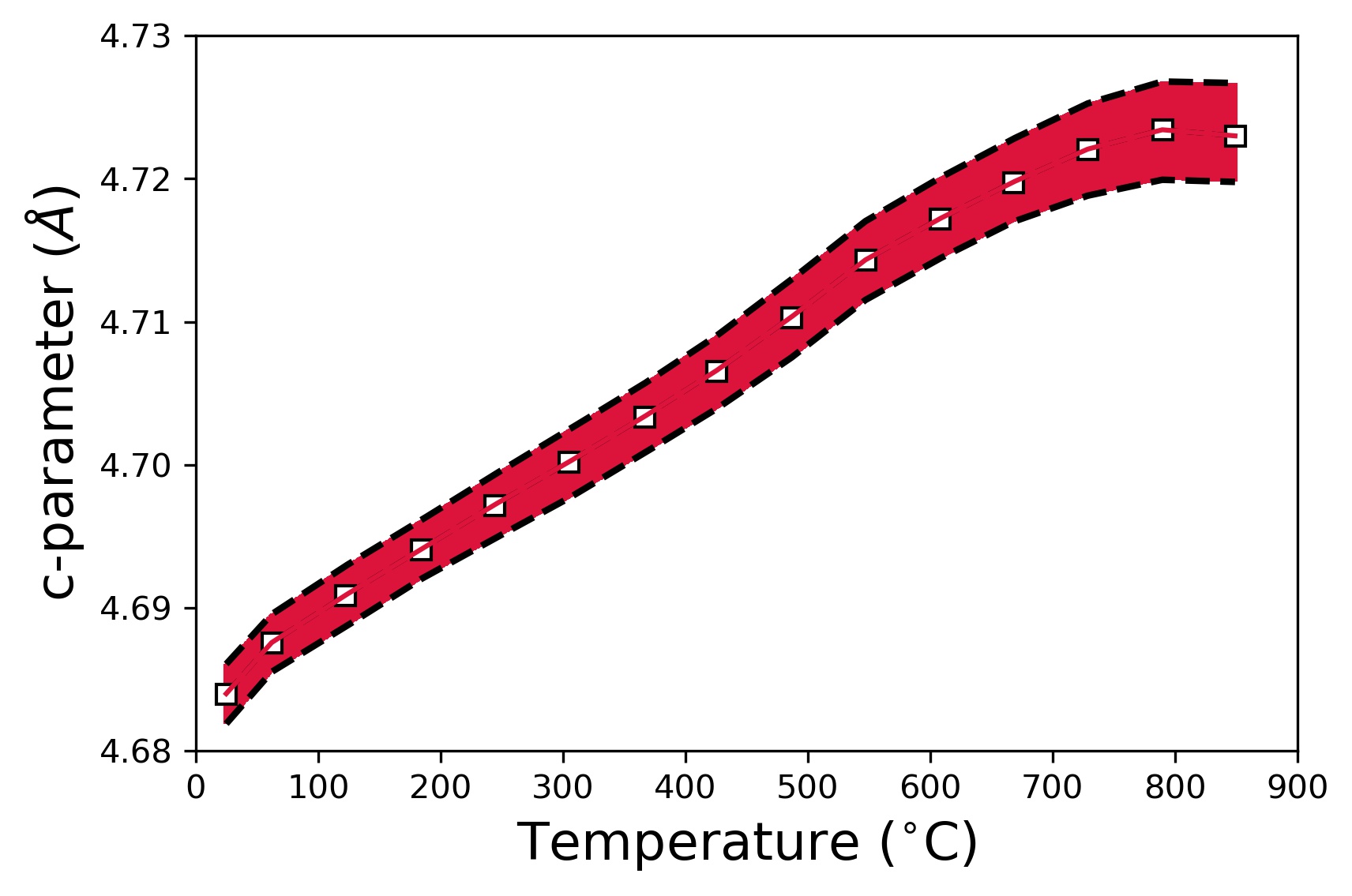}} \\
		
		\caption{The (a)-(b) a-parameter and (c)-(d) c-parameter are plotted as a function of temperature for both samples, where the square glyphs represent the mean values, and the colored bands represent the 10th and 90th percentile. The a- and c-parameters expanded with respect to temperature as expected.}
		\label{fig:lat_parm_all}
	\end{figure}

	Figure \ref{fig:lat_parm_all} shows the lattice parameters as a function of temperature. The white squares represent the average value for the ensemble of grains at a given temperature while the colored bands span the 10\textsuperscript{th} through 90\textsuperscript{th} percentile of the spread. The lattice parameters for each grain were calculated at each temperature by taking the magnitude of the deformed lattice vectors. Then, the average lattice parameters for the entire ensemble of grains is used as the lattice parameters of the material at that temperature. As expected, the a-parameter and c-parameter expand monotonically with increasing temperature. To confirm that reasonable results were being obtained, the f{}f-HEDM data was reduced to 1D profiles in 2$\theta$ by summing over images in $\omega$ then over $\eta$ to create a representative powder pattern, and the lattice parameters were extracted. The lattice parameters calculated using this method, and those from the aforementioned f{}f-HEDM method were in good agreement (see \ref{app:powder}).

\begin{figure}[b!]
		\centering
		\subfloat[Sample 1]{\includegraphics[width=0.45\textwidth]{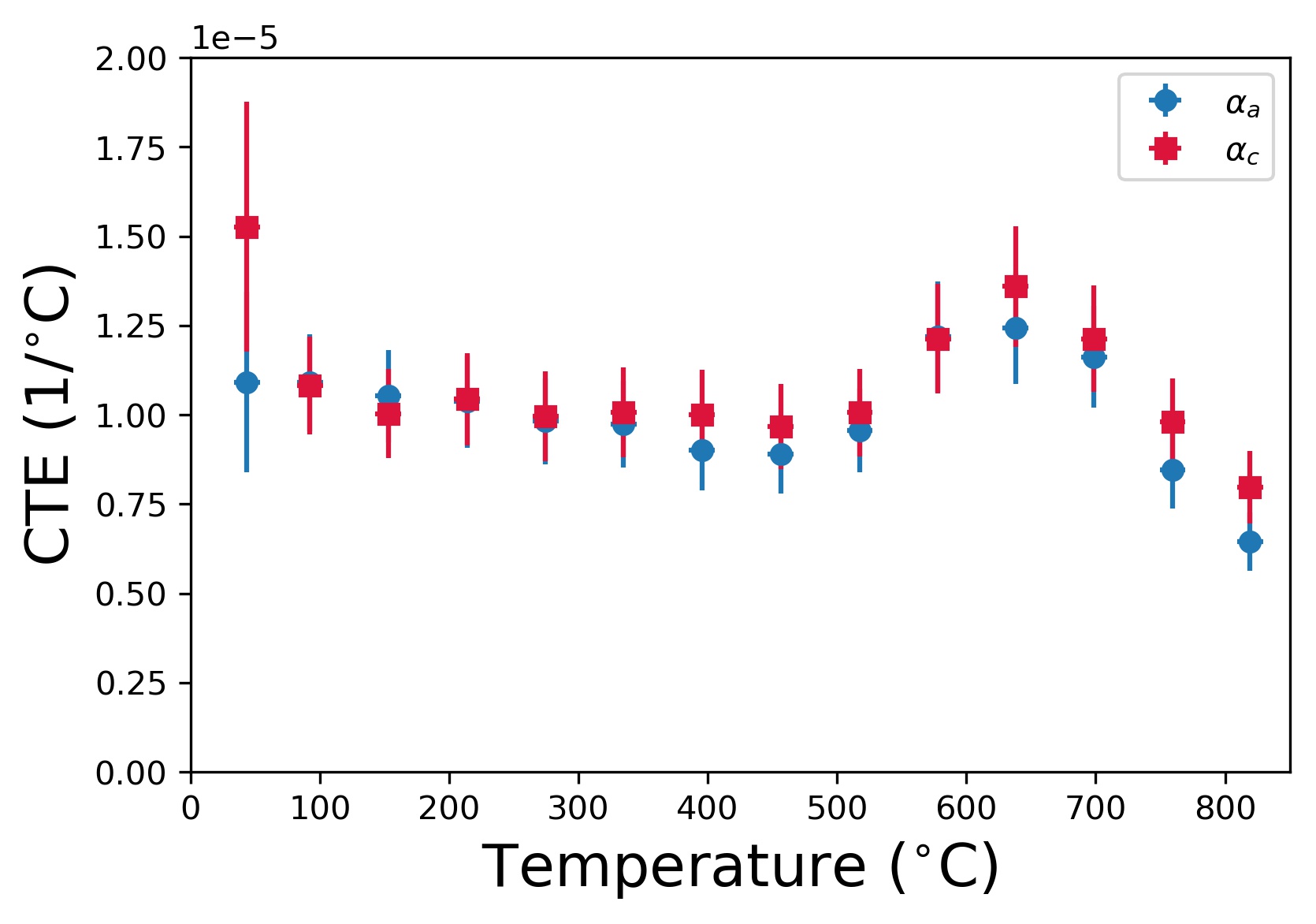}}\hspace{5pt}
		\subfloat[Sample 2]{\includegraphics[width=0.45\textwidth]{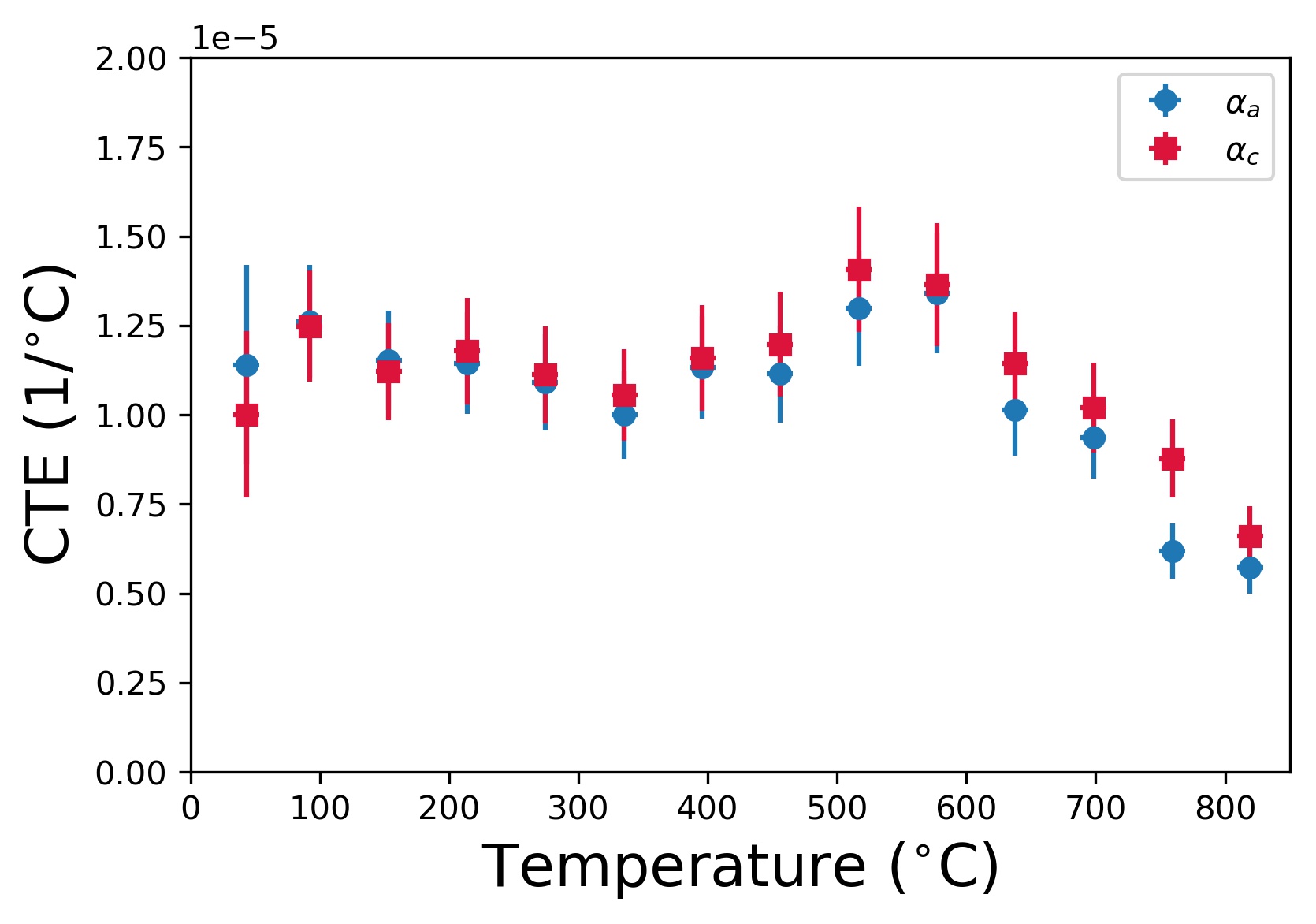}}\\
		\subfloat[]{\includegraphics[width=0.45\textwidth]{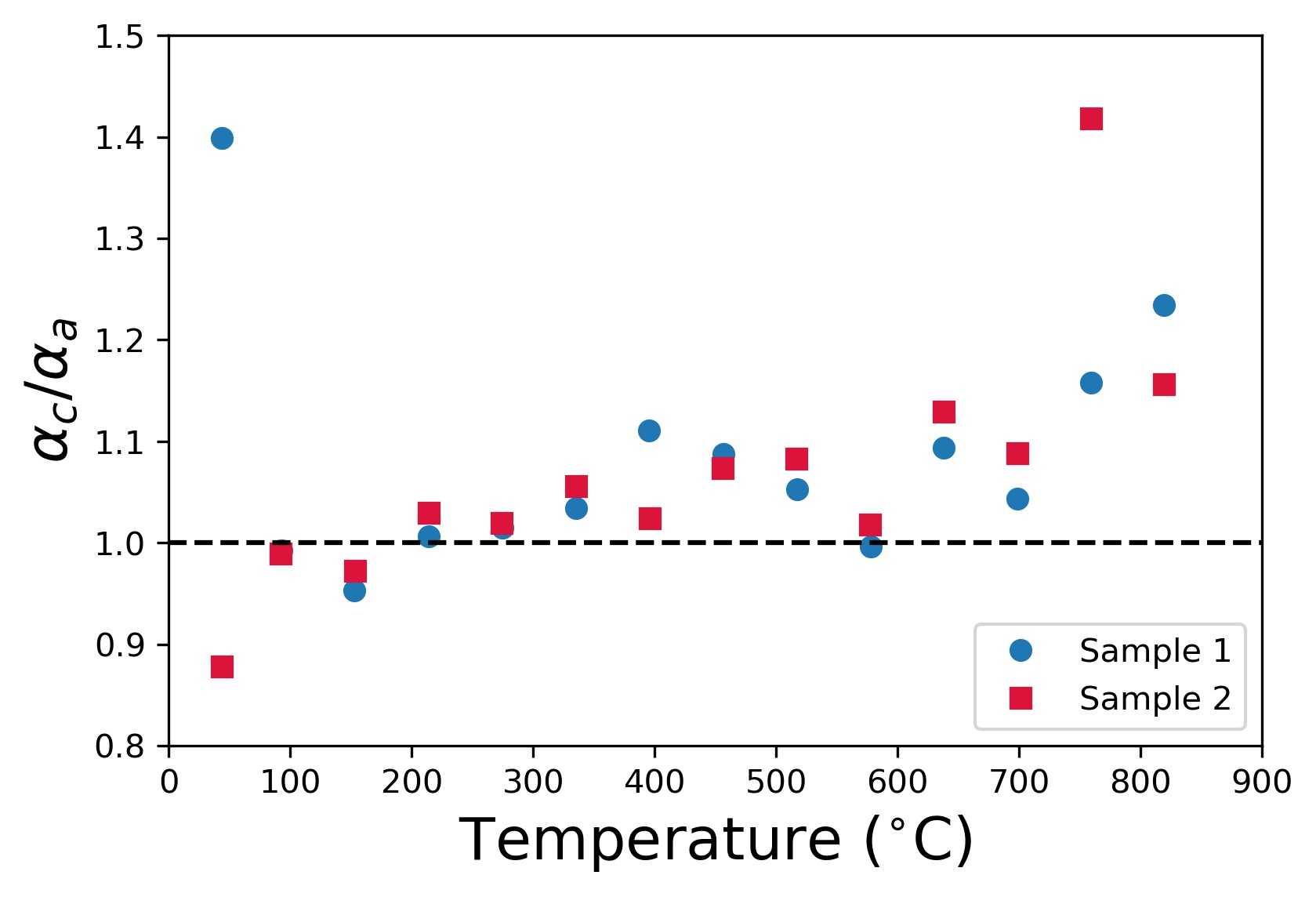}}
		\caption{(a)-(b) Both $\alpha_a$ and $\alpha_c$ decrease, then increase, then decrease again with temperature as the samples are heated. Error bounds are given from a Monte Carlo calculation. (c) The ratio of $\alpha_c$ to $\alpha_a$ increases as a function of T. The ratio is less than one at low temperature and is greater than one at high temperature. This variation helps to explain some of the inconsistency in the ratios reported in the literature. The CTEs are anisotropic (with $\alpha_c < \alpha_a$) at RT, reasonably isotropic from 150 to 600~$\degree$C, and anisotropic (with $\alpha_c > \alpha_a$) above this.}
		\label{fig:CTEs}
	\end{figure}

	The temperature dependent directional CTEs are shown for both samples in Figures. \ref{fig:CTEs}a and \ref{fig:CTEs}b respectively, where the bars represent the error bounds calculated from a Monte Carlo calculation, while Figure \ref{fig:CTEs}c shows the the relative CTE ratio, $\alpha_c$ over $\alpha_a$. This value increases with respect to temperature. The CTEs were calculated according to Eq. \ref{eq:alphaT}, and the ratios of $\alpha_a$ and $\alpha_c$ were then calculated from this data. Although the directional CTEs exhibited minor variations, in general, both follow similar trends with a peak near 600~$\degree$C. The ratio of $\alpha_c$ to $\alpha_a$ increased monotonically with heating, with a crossover in ratio from less than one to greater than one occurring at $\sim$170~$\degree$C as shown in Figure \ref{fig:CTEs}c. Additionally, the spread in lattice parameters broadens slightly with increasing temperature.
	
\subsection{Ti-7Al grain strain and stress evolution}	
\begin{figure}[h]
		\centering
		\subfloat{\includegraphics[height=0.55\textwidth]{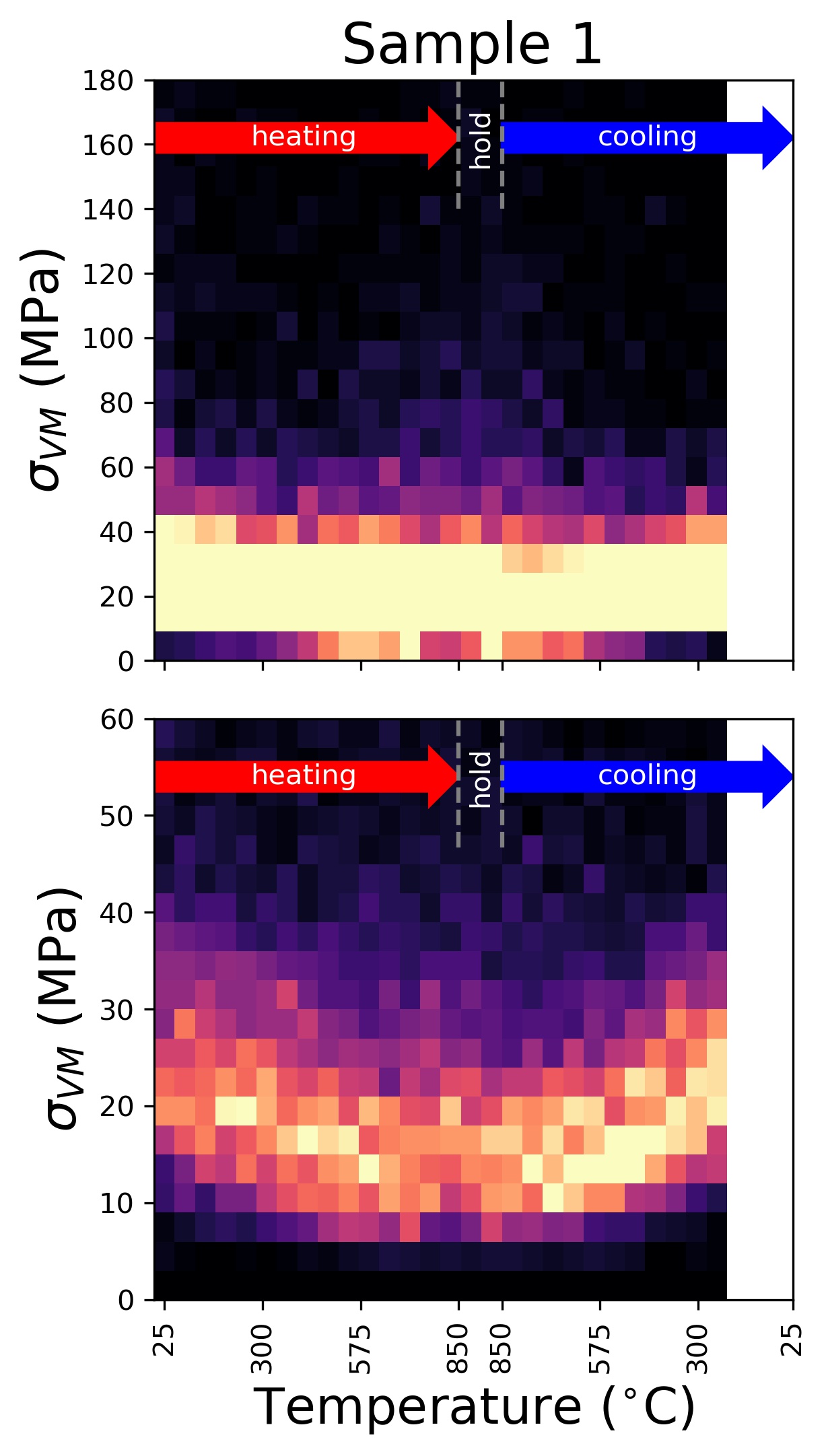}}\hspace{5pt}
		\subfloat{\includegraphics[height=0.55\textwidth]{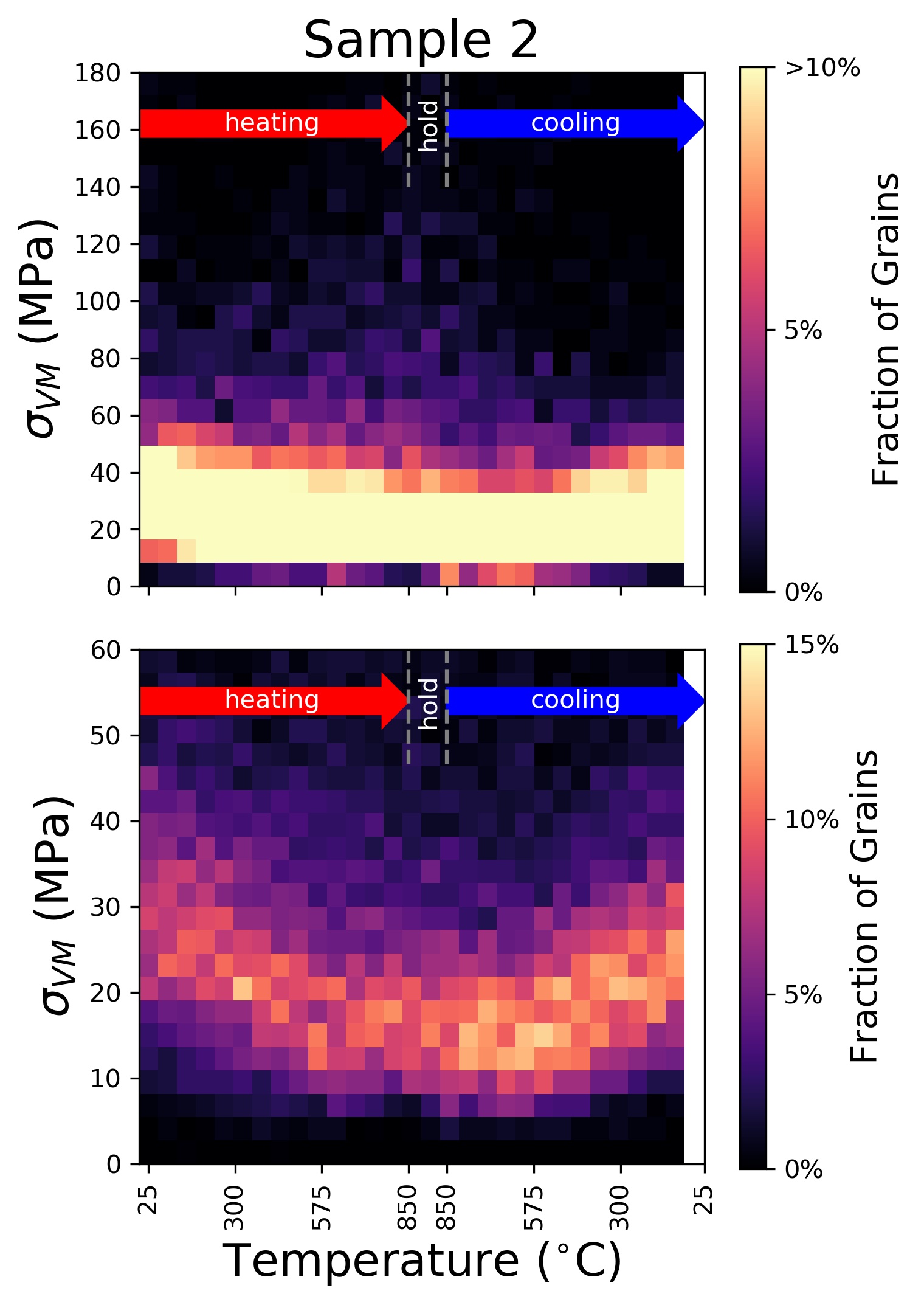}} \\

		\caption{The distribution of von Mises stress is plotted at each temperature where the top and bottom graphs have the same data but are scaled dif{}ferently to highlight key features. The larger stress range (top) shows an increase at the high end of the distribution at high temperature. The smaller stress range (bottom) shows that the von Mises stresses are lower after the thermal cycle, and the distributions become tighter. The white space on the right is where the x-ray beam was lost before the experiment reached room temperature.} \label{fig:full_thermal_cycle}		
	\end{figure}

	Von Mises stress is a commonly used measure of deviatoric stress for predicting plastic deformation in a material, and in the most basic sense, the higher the von Mises stress ($\sigma_{VM}$), the more likely a material is to yield. In order to examine the likelihood of possible micro-plastic events, the distribution of $\sigma_{VM}$ was plotted in Figure \ref{fig:full_thermal_cycle} for each temperature in the thermal cycle. The $\sigma_{VM}$ decreases until $\sim$700~$\degree$C when the spread of the distributions start to increase. As the samples cool, the mean and spread of the distribution of $\sigma_{VM}$ decreases until the sample reaches $\sim$700~$\degree$C when they start to increase again until room temperature where both the final mean and spread of the distributions are less than the initial values.

\begin{figure}[h!]
	\centering
	\subfloat[Sample 1]{\includegraphics[width=0.27\textwidth]{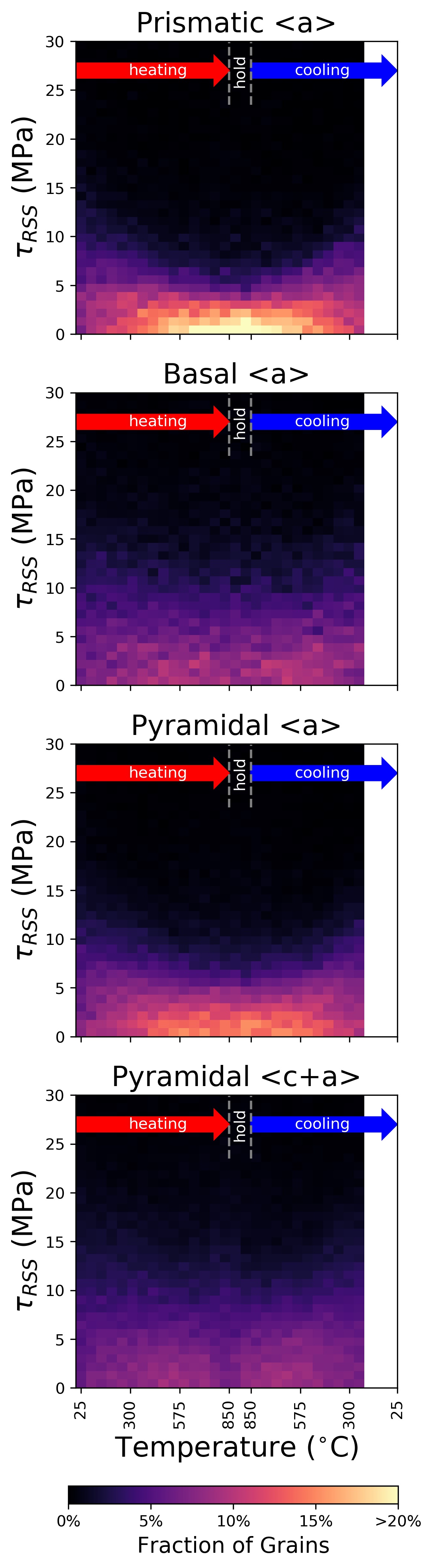}}\hspace{5pt}
	\subfloat[Sample 2]{\includegraphics[width=0.27\textwidth]{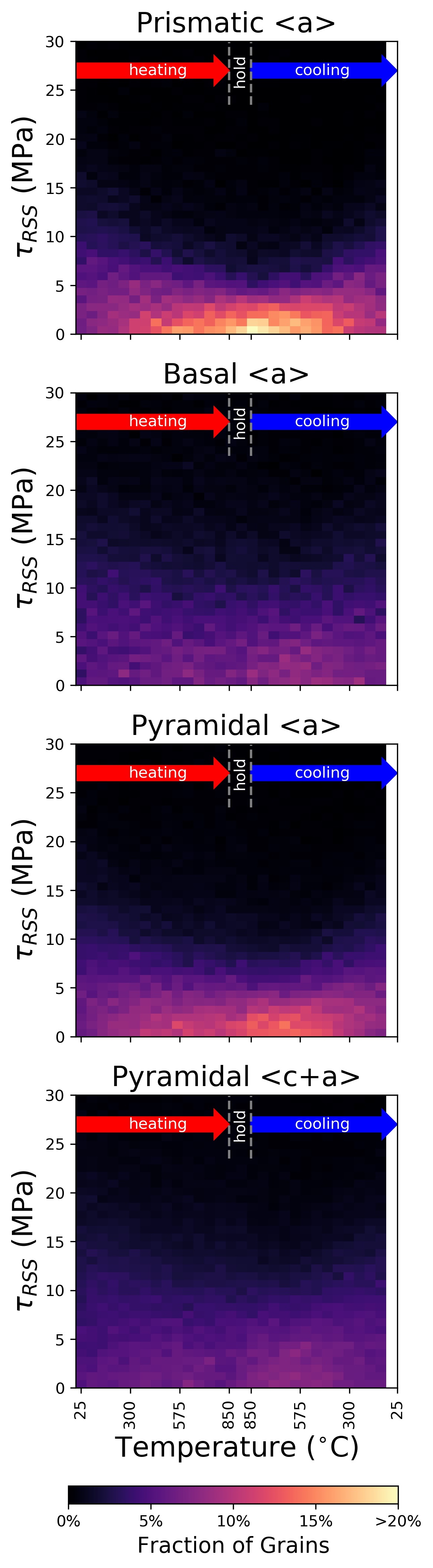}}\\

	\caption{The resolved shear stresses (RSS) for the prismatic $\langle$a$\rangle$, basal $\langle$a$\rangle$, pyramidal $\langle$a$\rangle$, and pyramidal $\langle$c+a$\rangle$, were calculated for (a) sample 1 and (b) sample 2 as function of temperature from the measured strain tensor and temperature dependent single crystal elastic modulus. The RSS values drop with temperature until $\sim$700~$\degree$C when the upper tail of the distributions for the pyramidal $\langle$c+a$\rangle$ RSS values starts to increase. The white space on the right is where the x-ray beam was lost before the experiment reached room temperature.}
	\label{fig:RSS}
\end{figure}

	Motivated by the possibility that certain grains may have experienced micro-plastic flow, the measured strains were used to calculate the elastic strain (Eq. \ref{eq:total lattice strain}) and subsequently the stress in each grain via temperature-dependent moduli from Fisher \textit{et al.} \cite{Fisher1964} (see \ref{app:CRSS}). The RSS values for all four slip systems were calculated by projecting the stress onto each individual system. Figure \ref{fig:RSS} shows that the RSS on all of the slip systems tends to decrease as a function of temperature. However, the mean values for the pyramidal $\langle$c+a$\rangle$ systems decrease until $\sim$700~$\degree$C when they start to increase slightly, similar to the $\sigma_{VM}$ distributions. Additionally, the distributions for the prismatic $\langle$a$\rangle$ and the pyramidal $\langle$a$\rangle$ systems are tighter than the distributions for the basal $\langle$a$\rangle$ and the pyramidal $\langle$c+a$\rangle$ systems.

\section{Discussion} \label{sec:discussion}
	Thus far, we have shown that f{}f-HEDM data can be used to measure not only the anisotropic thermal expansion in Ti-7Al, but also the ef{}fects of the grain-interactions caused by the constraints of a polycrystalline aggregate. This section further discusses these results and their significance.

\subsection{Comparison of CTEs with literature}
	As mentioned, the CTEs for titanium reported in literature (Table \ref{tab:Ti_CTE}) vary from a low of 5.6~$\times$~10\textsuperscript{-6}/$\degree$C to a high of 13.6~$\times$~10\textsuperscript{-6}/$\degree$C and show little agreement even for the ratio of $\alpha_a$ to $\alpha_c$. Based on the results of this study (Fig. \ref{fig:CTEs}), the temperature dependence of the CTEs may explain the observed spread of values and ratios. Most reported values in the literature do not take temperature dependence into account, whereas we find that the CTE ratio is less than one at low temperatures and greater than one at high temperatures, therefore the temperature range in which the CTEs are measured is critical.

	For example, Pawar and Deshpande \cite{Pawar1968} and McHargue and Hammond \cite{Mchargue1953} measured the CTEs up to 155~$\degree$C and 225~$\degree$C respectively, and report values where $\alpha_c < \alpha_a$. This is within the temperature range where the present work shows that the $\alpha_c$ to $ \alpha_a$ ratio is less than one. In contrast, Berry and Raynor \cite{Berry1953} and Spreadborough and Christian \cite{Spreadborough1959} measured the CTE up to 700~$\degree$C and 600~$\degree$C, and report values where $\alpha_c > \alpha_a$. In this temperature range, the present work confirms that the ratio of $\alpha_c$ to $ \alpha_a$ is greater than one. This also agrees with the results from Zheng et al. \cite{Zheng2019}, who showed a crossover point around 690~$\degree$C. However, these measurements were made in commercial purity titanium and were extracted from a simulation, and the experimental CTEs were not shown. The crossover point found here at 170~$\degree$C for the Ti-7Al alloy is lower than what was found in commercial purity Ti \cite{Zheng2019} but follows the same trend.

\begin{figure}[h!]
		\centering
		\includegraphics[width=0.35\textwidth]{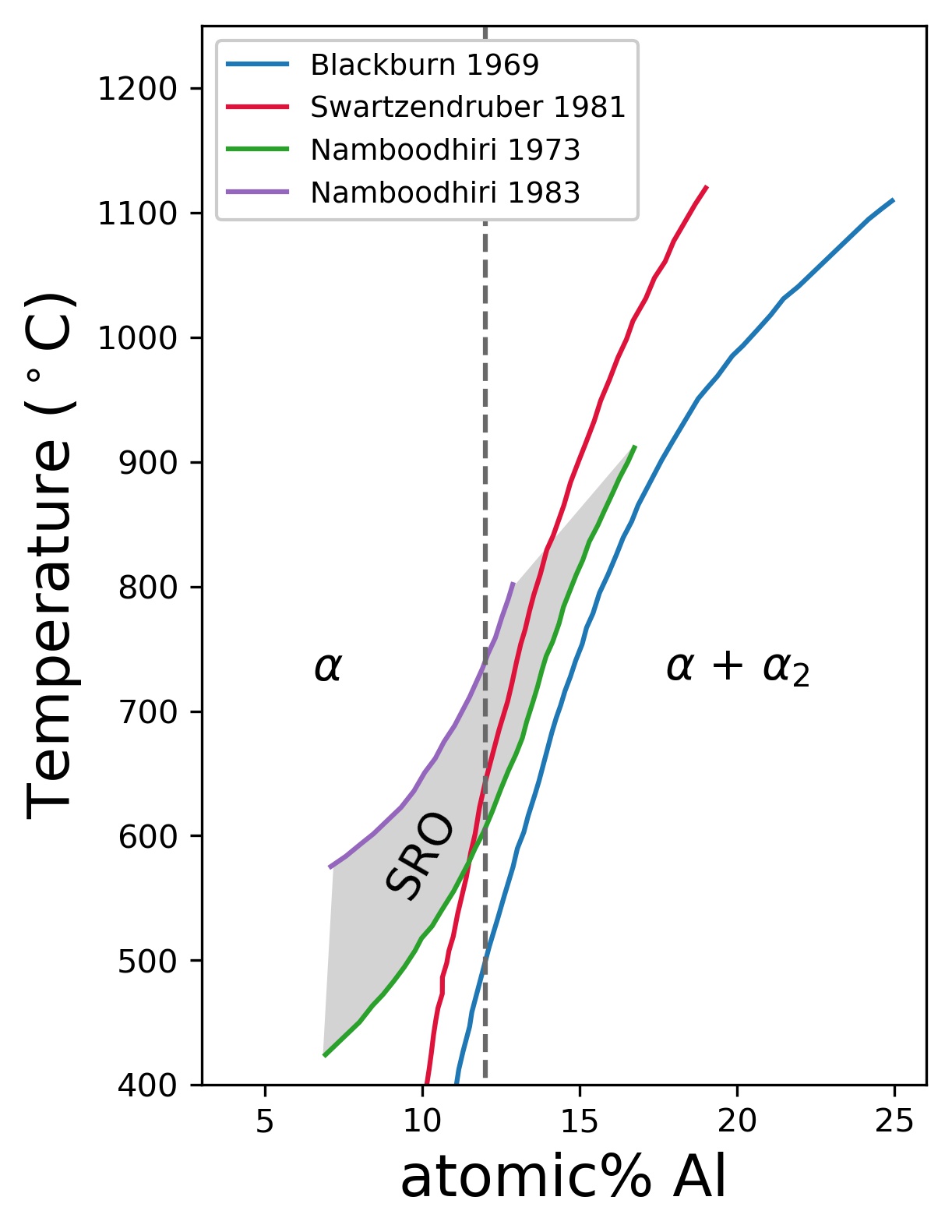}
		\caption{Ti-rich section of the Ti-Al phase diagram adapted from  \cite{Blackburn1969,Swartzendruber1981,Namboodhiri1973,Namboodhiri1983} where 12 at\% Al is equal to 7 wt\%.}
		\label{fig:phase_diagram}
	\end{figure}

	In addition, it is important to note that with 7 wt\% Al (12 at\%) , the Ti-7Al material falls within the two-phase region of the Ti-Al phase diagram (Figure \ref{fig:phase_diagram}). This can explain why the CTE values for the two axes vary systematically with temperature, with three distinct sections in the curve (Fig. \ref{fig:CTEs}a-b). The first section is the decreasing CTE of the Ti-7Al material with both the $\alpha$ and $\alpha_2$ phases present, the second region is the increasing CTE where the $\alpha_2$ nanoprecipitates are dissolving into the material, and the third region is the Ti-7Al material with only the $\alpha$ phase. However, the boundary between the $\alpha$ and $\alpha + \alpha_2$ regions is not well defined as can be seen in Figure \ref{fig:phase_diagram}. For this composition, the Ti\textsubscript{3}Al and SRO regions dissolve somewhere between 500 $\degree$C and 700 $\degree$C, but the exact temperature is unknown. The large temperature range can explain the discrepancy between samples 1 and 2. It is expected that if the material was to be heated with a more gradual ramp rate, the increase in CTE would be sharper and more distinct as the $\alpha_2$ should dissolve at a single temperature.
	
\subsection{Micromechanical Response}
	At 700~$\degree$C, the RSS values for the basal, prismatic, and pyramidal $\langle$a$\rangle$ slip system families continue to decrease whereas the pyramidal $\langle$c+a$\rangle$ slip systems increase markedly. The pyramidal $\langle$c+a$\rangle$ slip system is harder than the other three \cite{Williams2002,Pagan2018}, so by 700~$\degree$C, the other three are soft enough that they experience slip, causing new grain interactions and a stress build up on the pyramidal $\langle$c+a$\rangle$ systems which still have some strength. This build up of stress on the pyramidal $\langle$c+a$\rangle$ slip systems is enough that it causes an increase in the upper tail of the $\sigma_{VM}$ distribution at high temperature. As the material is cooled back down to room temperature, the grains contract, and the internal stresses follow the reverse path of the heating. 
		
\begin{figure}[t!]
		\centering
		\subfloat[]{\includegraphics[width=0.48\textwidth]{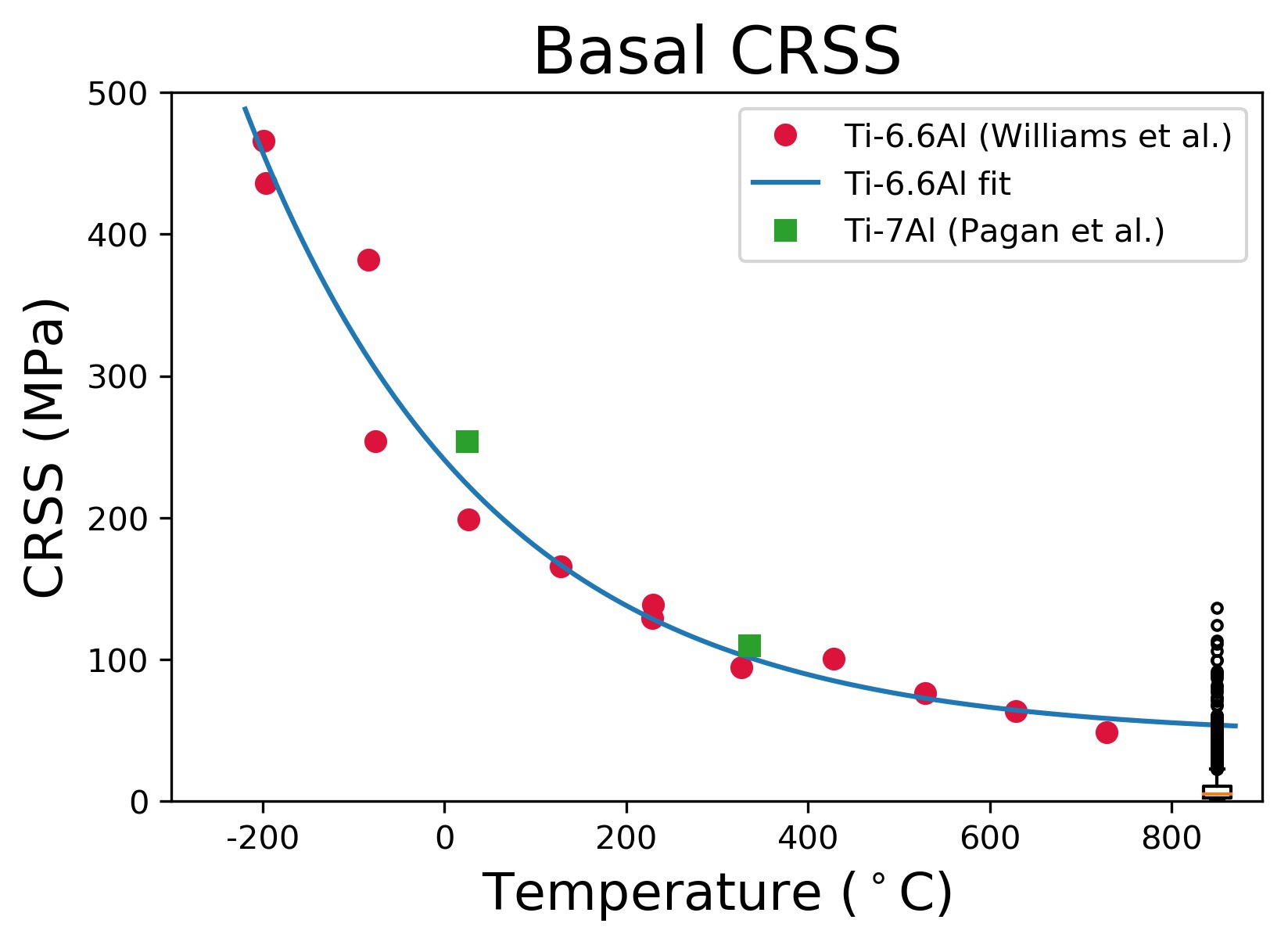}}\hspace{5pt}
		\subfloat[]{\includegraphics[width=0.48\textwidth]{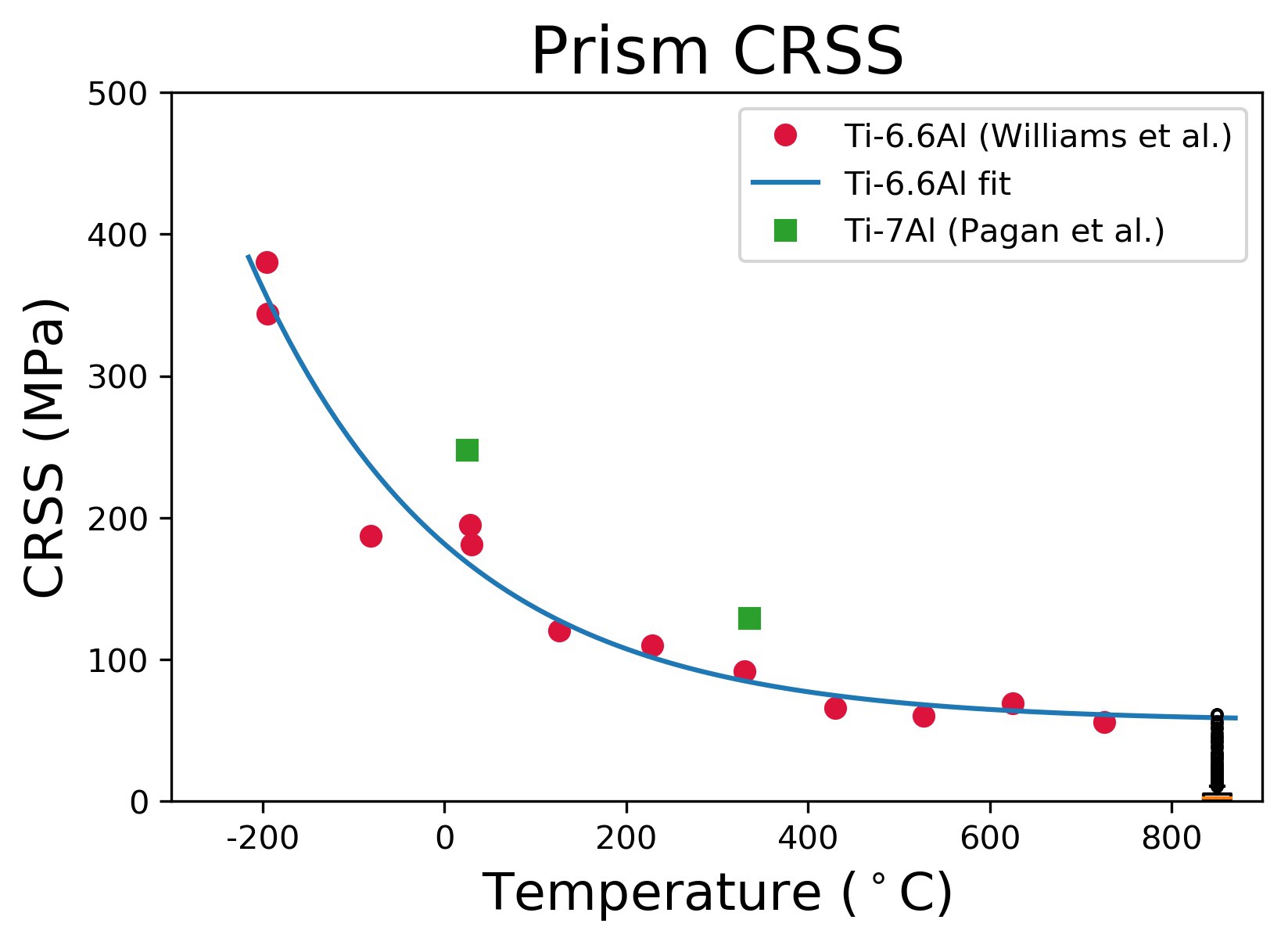}}
		\caption{(a) Plot of the basal CRSS as a function of temperature with a boxplot of the RSS values superimposed to show that the outlier RSS values easily exceed the CRSS at high temperature. (b) Plot of the prismatic CRSS as a function of temperature with a boxplot of the RSS values superimposed to show that even the outlier RSS values do not significantly exceed the CRSS at high temperature. CRSS values taken from Williams \textit{et al.} \cite{Williams2002} and Pagan \textit{et al.} \cite{Pagan2017, Pagan2018}}.
		\label{fig:CRSS_with_Boxplot}
	\end{figure}

	However, what is relevant here is whether the RSS values exceed their associated CRSS. Williams \textit{et al.} \cite{Williams2002} measured CRSS values for Ti-6.6Al prismatic and basal slip systems in compression. The values for Ti-7Al determined by Pagan \textit{et al.} \cite{Pagan2018} at room temperature and at 355~$\degree$C follow the trend line fit to the Ti-6.6Al data that shows an exponential decay in the slip system strength as a function of temperature. Figure \ref{fig:CRSS_with_Boxplot} shows the measured CRSS values along with the exponential fit as well as a boxplot for RSS values at the highest temperature.  In the case of the prismatic slip system, the RSS values do not significantly exceed the CRSS, but for the basal system, there are many grains in the upper tail whose basal RSS exceeds the CRSS indicating the possibility for slip to occur. 
	
\begin{figure}
		\centering
		\subfloat{\includegraphics[height=0.29\textheight]{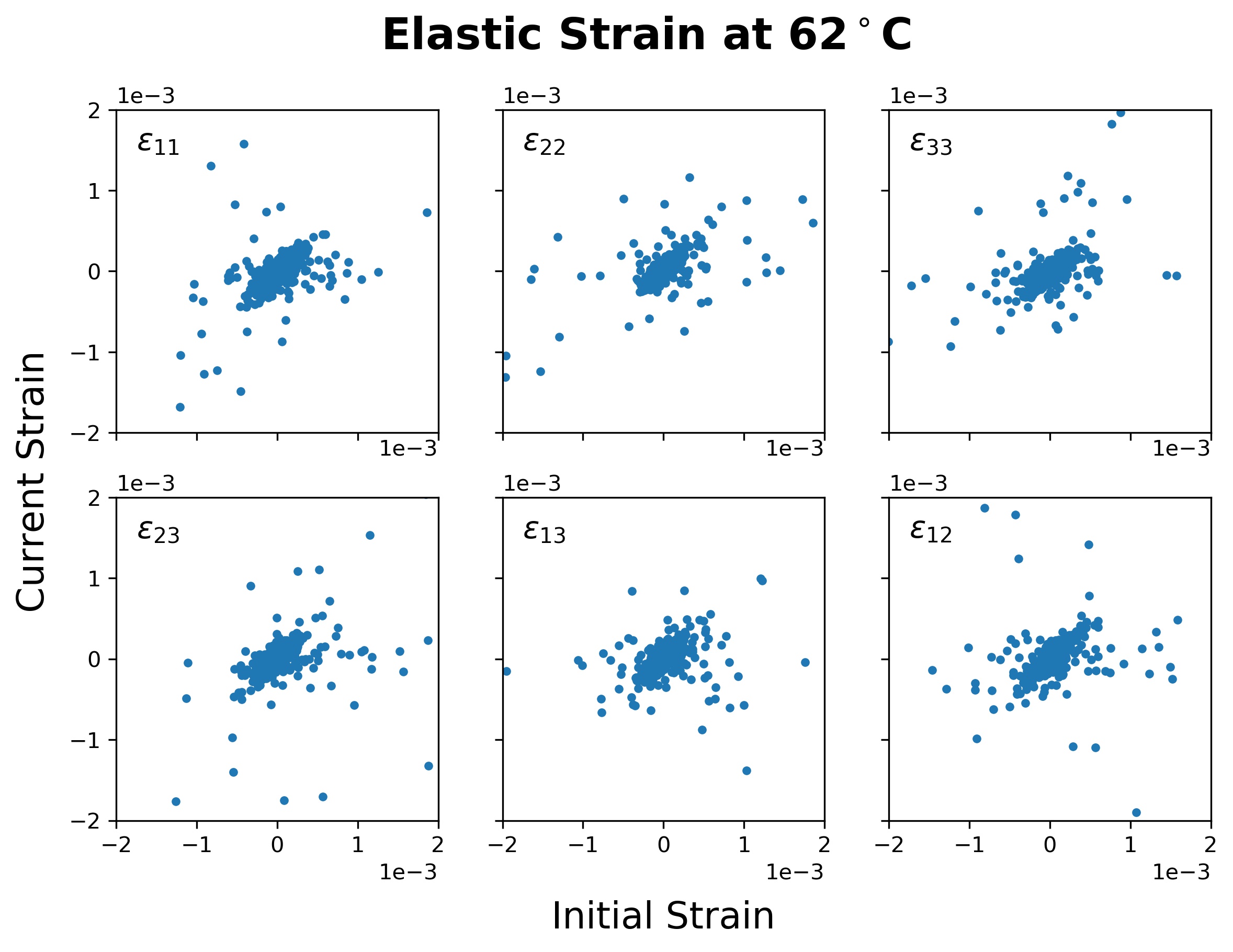}} \\
		\subfloat{\includegraphics[height=0.29\textheight]{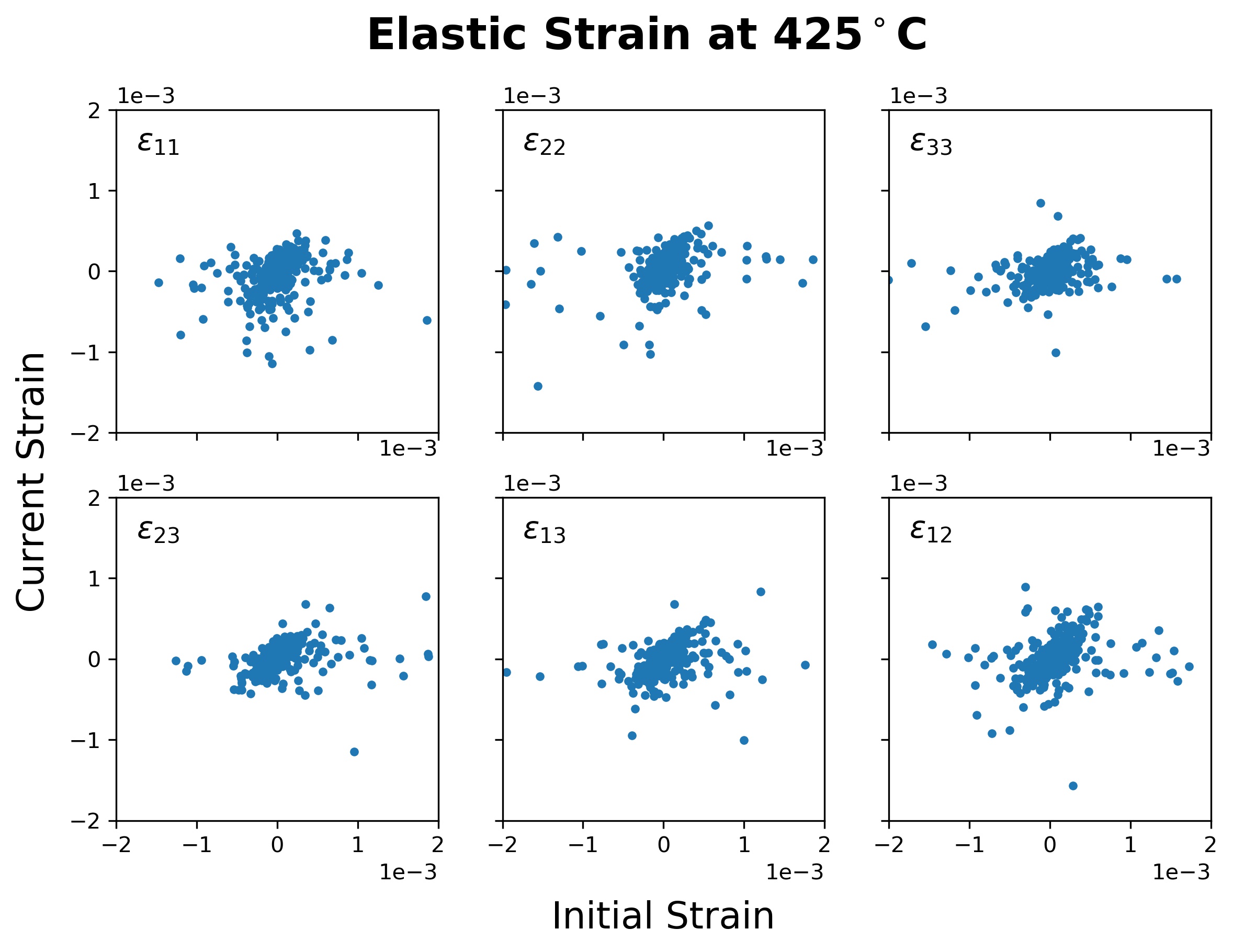}} \\
		\subfloat{\includegraphics[height=0.29\textheight]{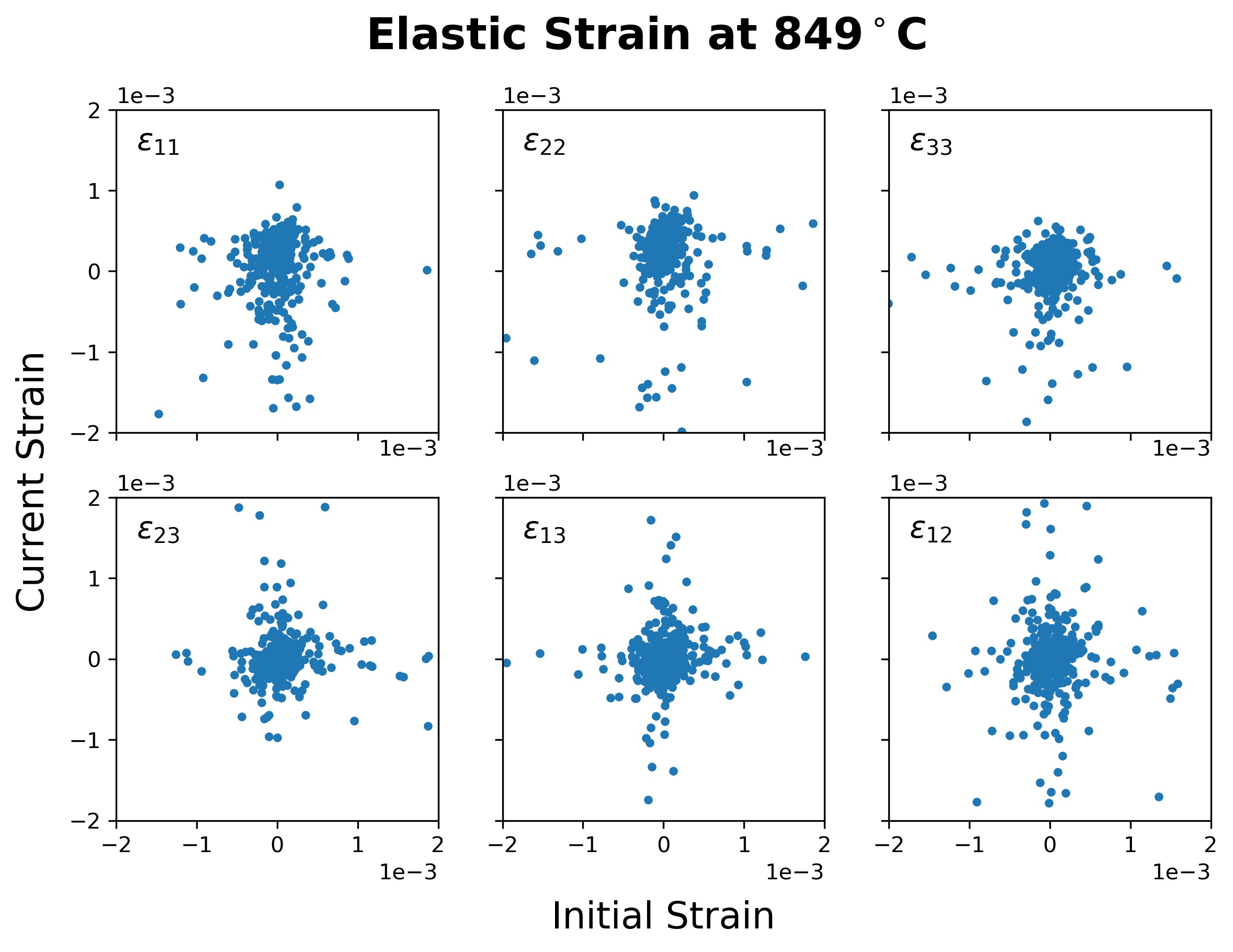}}
		\caption{The elastic strain is plotted against the initial elastic strain for each grain at a given temperature. As the sample heats up, the correlation between the initial and current elastic strain weakens and becomes negligible at the peak temperature.}
		\label{fig:elastic_change}
	\end{figure}

	This is further supported by the change in the elastic strain state of each of the grains compared with their initial state shown in Figure \ref{fig:elastic_change}. A correlation can be seen between the initial elastic strain and the elastic strain at the first thermal step at 62 $\degree$C. By the time the temperature reaches 849 $\degree$C, there is no obvious correlation. In order to quantify this loss of correlation with respect to temperature, a Pearson correlation coef{}ficient was calculated at each temperature:
	\begin{equation}
		\rho_{x,y} = \frac{\mathrm{cov}(x,y)}{\sigma_x \sigma_y}
	\end{equation}
	where $\sigma$ is the standard deviation of $x$ and $y$ respectively, and $\mathrm{cov}$ is the covariance given by
	\begin{equation}
		\mathrm{cov}(x,y) = E \, [(x-E[x]) \, (y-E[y])]
	\end{equation}
	where $E[x]$ is the expected value of x. A Pearson correlation coeeficient value of one indicates complete correlation and a value of zero indicates no correlation

\begin{figure}[h!]
		\centering
		\subfloat{\includegraphics[height=0.43\textheight]{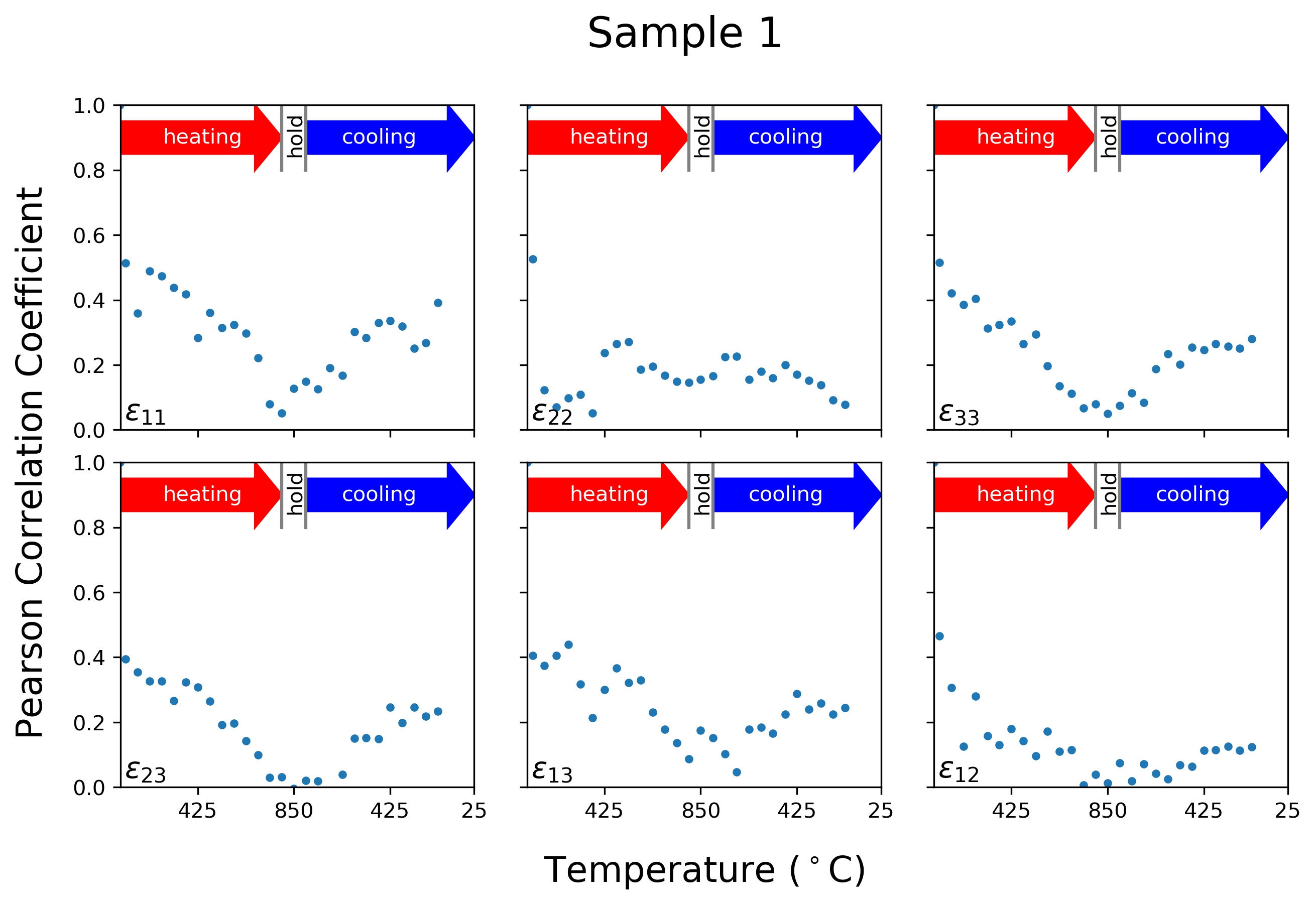}} \\
		\subfloat{\includegraphics[height=0.43\textheight]{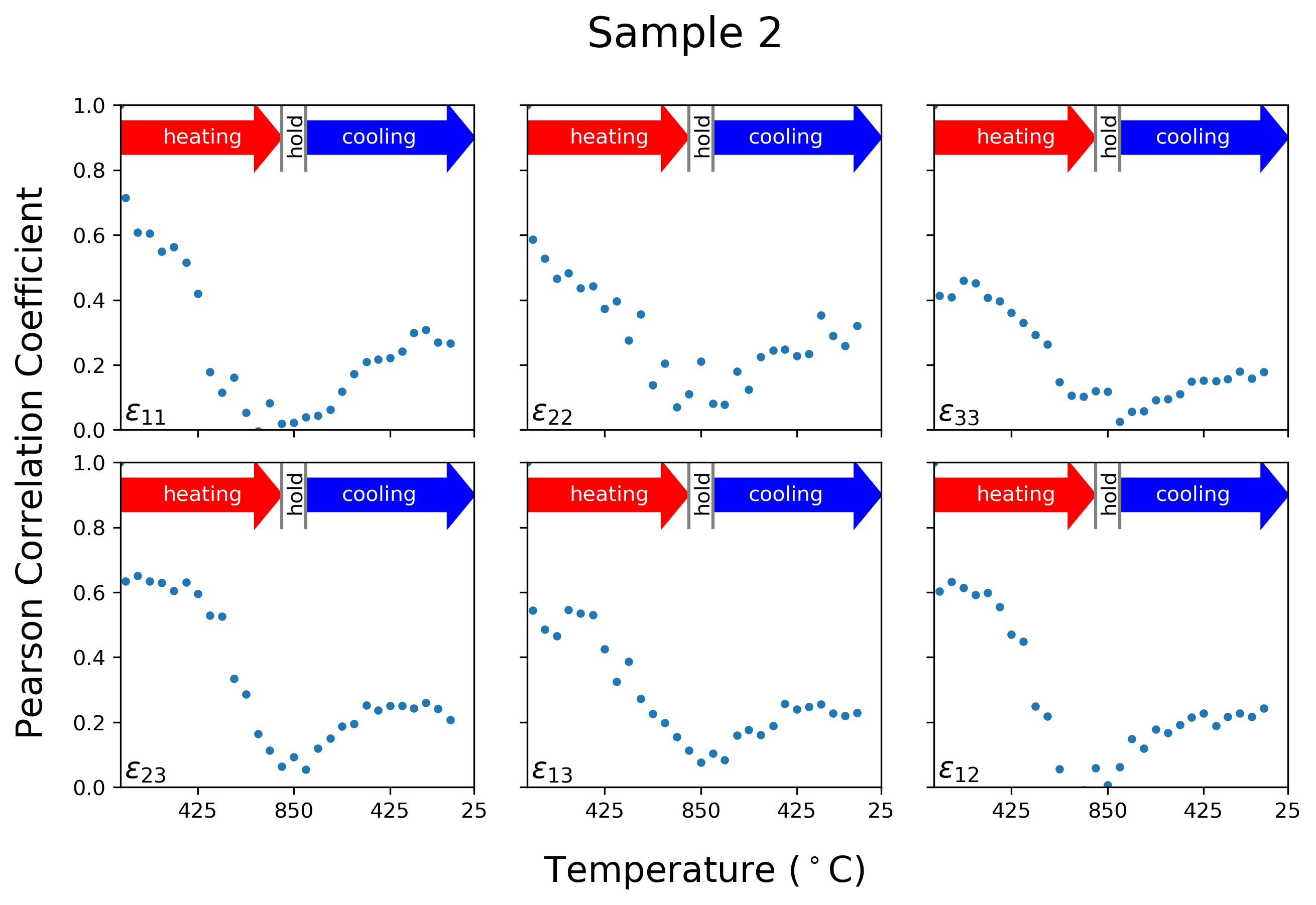}}
		\caption{The Pearson correlation coef{}ficient was calculated between the elastic strain state at a given temperature and the initial elastic strain state for each strain component.}
		\label{fig:pearson}
	\end{figure}

	Figure \ref{fig:pearson} shows that the correlation according to the calculated Pearson correlation coef{}ficient goes down as the temperature increases, reaching a minimum of about zero at the highest temperature, then the correlation increases again as the temperature drops. The strain values at the maximum temperature are effectively uncorrelated with the initial values, and yet a correlation is partially recovered upon cool-down, which indicates that the polycrystal retained some ``memory" of its initial state, albeit with local changes. This indicates that the thermal cycle was not purely thermoelastic, signifying the occurrence of annealing during the thermal cycle. Additionally, the lack of correlation at high temperature is presumably caused by the anisotropic thermal expansion overpowering the initial elastic strain state.
	
\section{Summary}

	In this work, f{}f-HEDM has been used to calculate the temperature dependent CTEs for hexagonal Ti-7Al with the result that the ratio of $\alpha_a$ to $\alpha_c$ changes from less than one to greater than one, resolving the discrepancies found for Ti thermal expansion data found in the literature. A sample dependent maximum was found to be between 500 and 650 $\degree$C which can be attributed to the dissolution of the Ti\textsubscript{3}Al. Additionally, the f{}f-HEDM measurements allow us to calculated confidence bounds for the CTEs. 
	
	It was also found that the RSS values for the three $\langle$a$\rangle$ slip system families decrease as a function of temperature while the RSS values for the pyramidal $\langle$c+a$\rangle$ slip systems decreases until around 700 $\degree$C when they start to increase. It is our understanding that the $\langle$a$\rangle$ slip system families are soft enough to allow slip to occur, but the pyramidal $\langle$c+a$\rangle$ is harder, so stress builds up on this slip system. Lastly, annealing was observed on the grain-scale during the thermal cycle as was indicated by the change in the grain-resolved elastic strain states.

	Future work on this data will focus on modeling this experiment with both FE and FFT-based crystal plasticity methods. This will allow us to gain a better understanding of what is occurring on the grain-scale and further tune the CTEs from those calculated in this work. This data has shown that there is more happening on the mesoscale than originally expected, and understanding the anisotropy of thermal expansion is important for the accuracy of micromechanical modeling work.

\section*{Acknowledgements}
	The authors would like to thank Adam Pilchak for providing the Ti-7Al material. REL would like to thank Yu-Feng Shen, Ziheng (Dino) Wu, and Linghan Zhang for their help in data collection and Benjamin Hulbert, David Dye, and Jerard Gordon for helpful discussions. This work is based upon research conducted at the Cornell High Energy Synchrotron Source (CHESS) which is supported by the National Science Foundation and the National Institutes of Health/National Institute of General Medical Sciences under NSF award DMR-1332208. The work is funded by the Air Force Of{}fice of Scientific Research under grant FA9550-16-1-0105.

\newpage
\begin{appendices}
\appendix

\section{Powder Dif{}fraction}\label{app:powder}

The f{}f-HEDM images were summed over the $\omega$ range to create a representative powder pattern. This was treated as a typical powder dif{}fraction pattern, and the lattice parameters were extracted. The lattice parameters calculated using this method, and those from the aforementioned f{}f-HEDM method were in good agreement.

\begin{figure}[h]
	\centering
	\subfloat[Sample 1]{\includegraphics[width=0.4\textwidth]{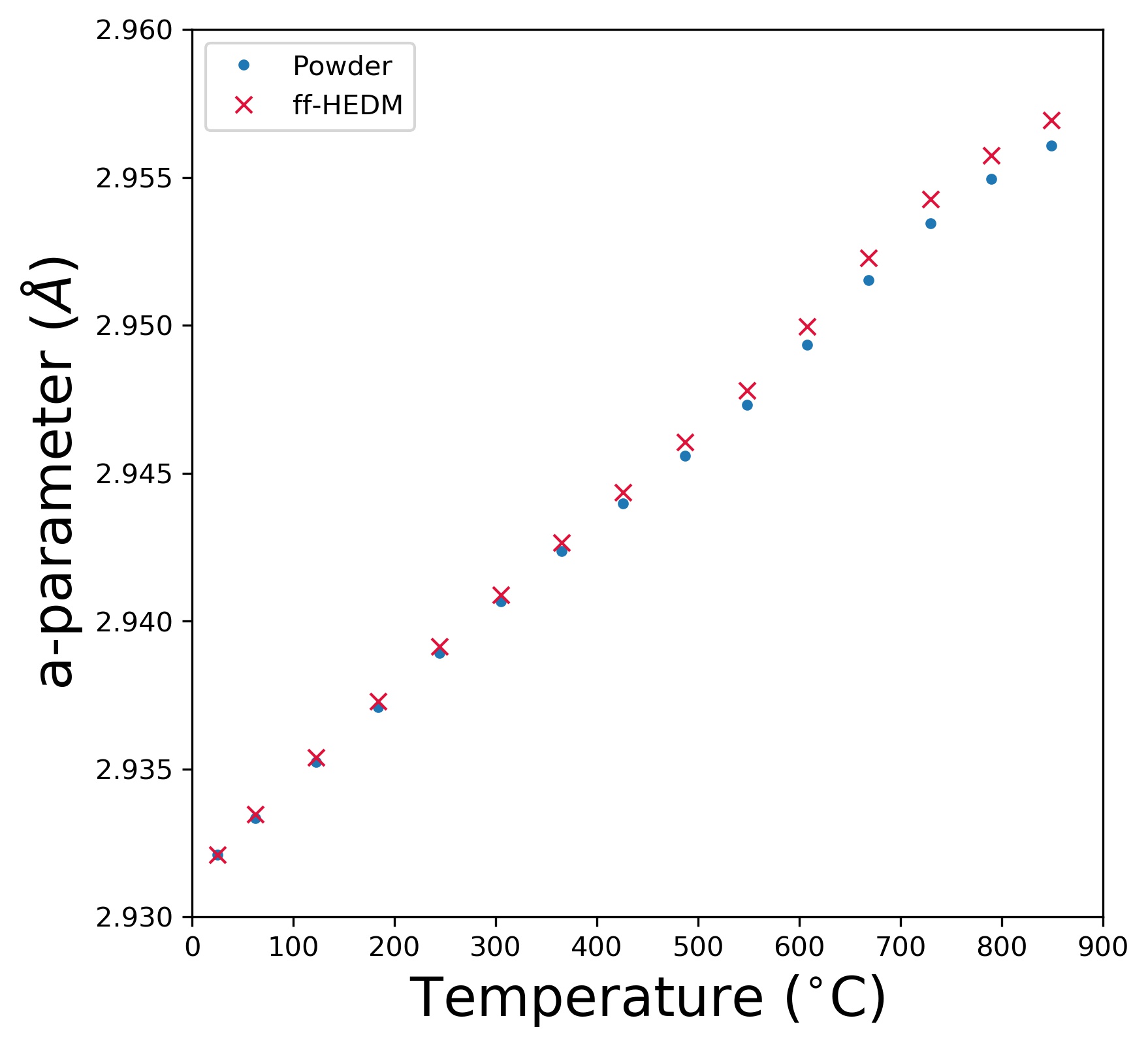}}\hspace{5pt}
	\subfloat[Sample 1]{\includegraphics[width=0.4\textwidth]{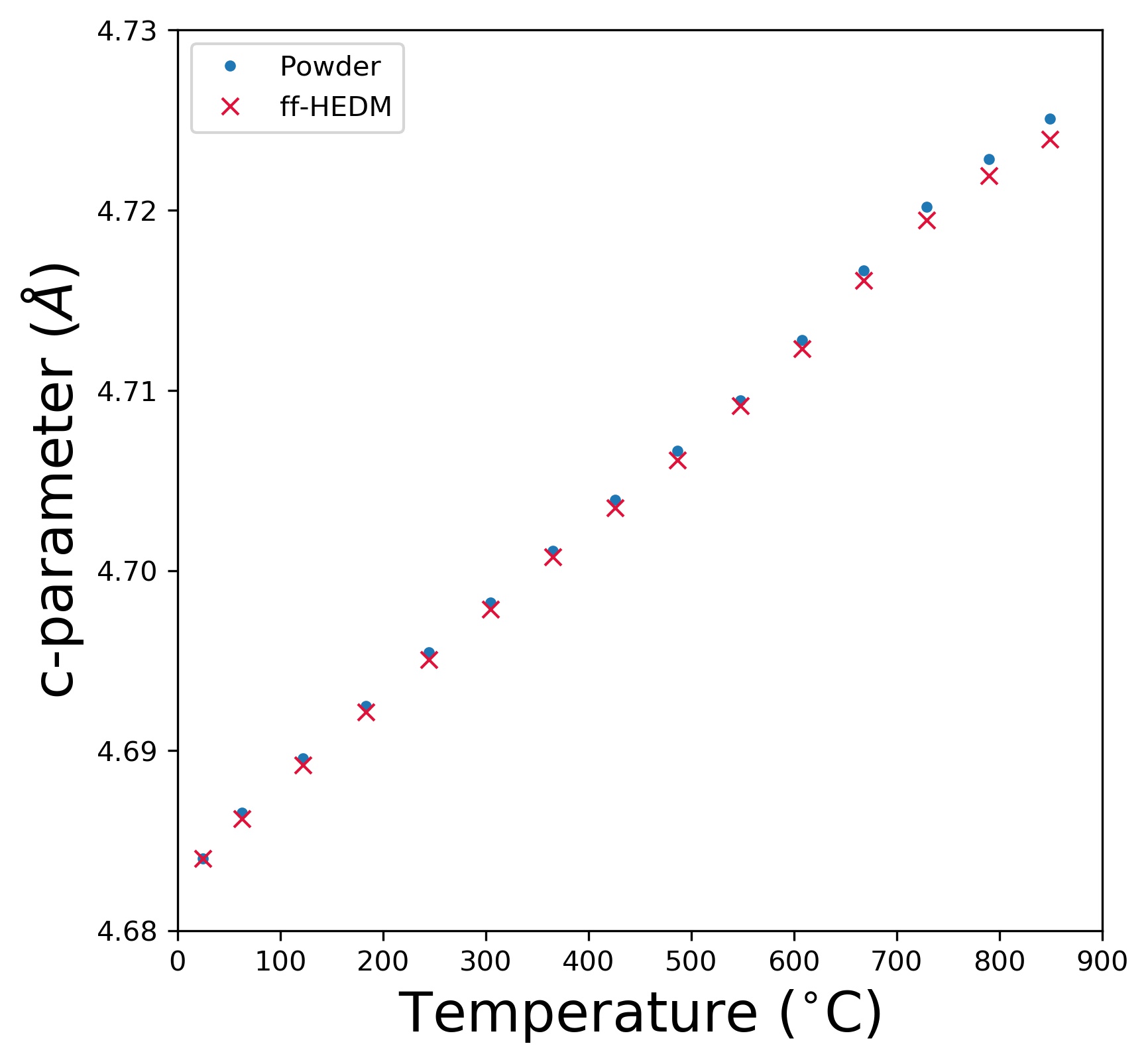}} \\
	\subfloat[Sample 2]{\includegraphics[width=0.4\textwidth]{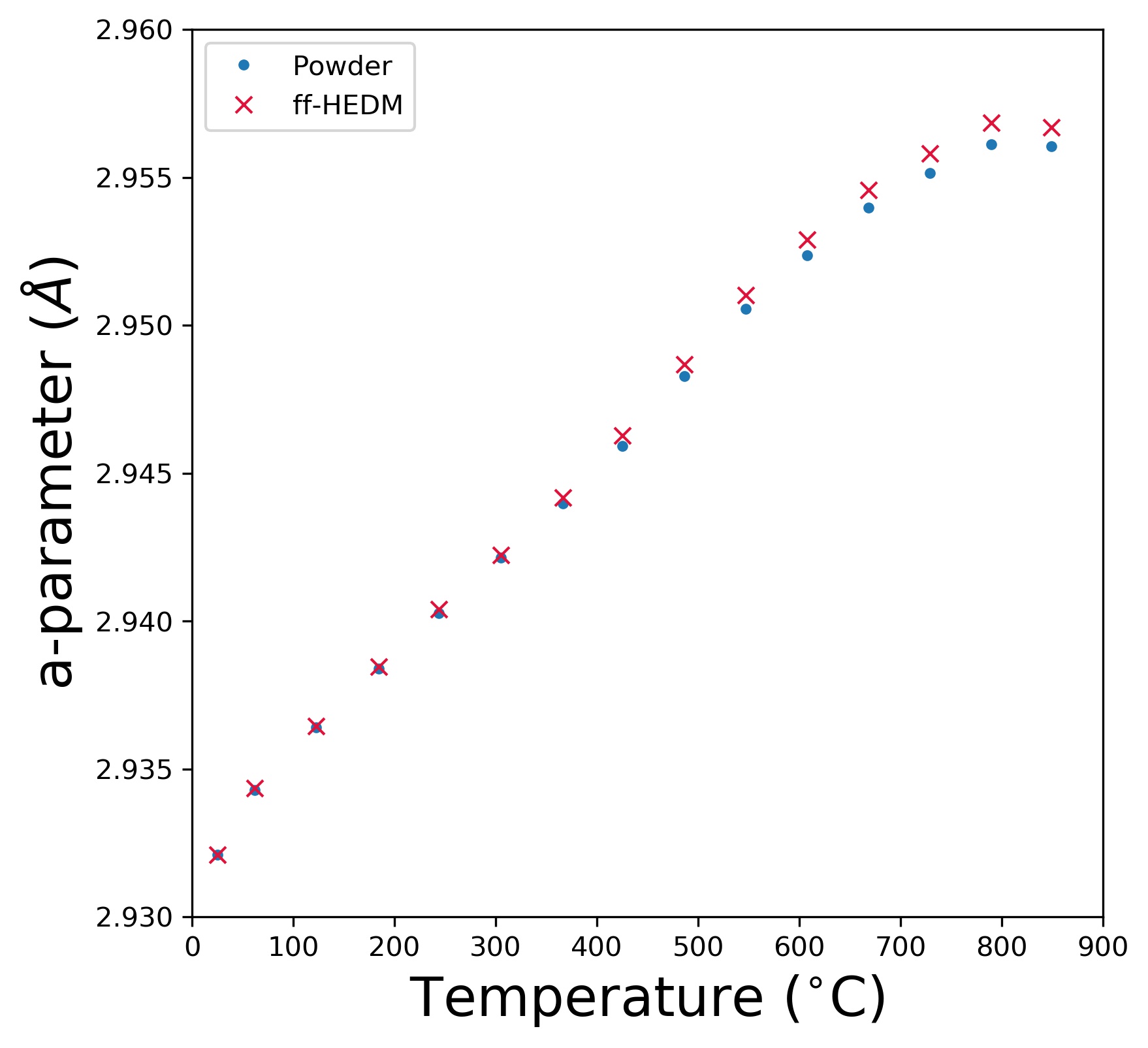}}\hspace{5pt}
	\subfloat[Sample 2]{\includegraphics[width=0.4\textwidth]{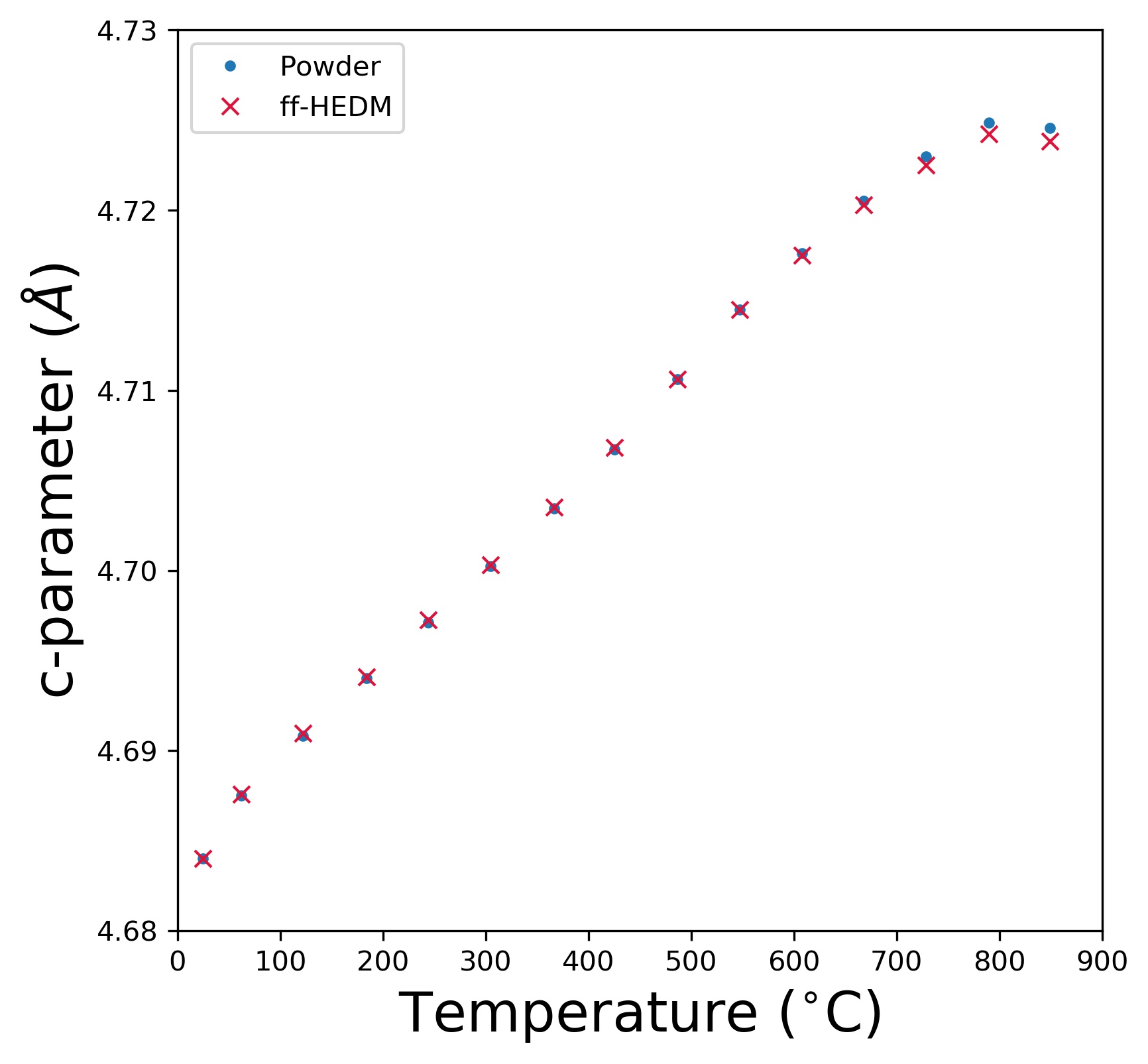}} 
	\caption{The f{}f-HEDM images were integrated across the entire rotation series, and the integrated images were treated as powder dif{}fraction patterns for both (a-b) sample 1 and (c-d) sample 2. The calculated lattice parameter matched those calculated from the f{}f-HEDM method.}
	\label{fig:powder}
\end{figure}

\newpage
\section{Calculating Resolved Shear Stresses}
\label{app:CRSS}
	When metals are placed under small loads, they deform non-permanently, or elastically, in such a way that the stress and strain are linearly proportional. In an anisotropic single crystal, this relationship can be calculated through the tensorial relationship
		\begin{equation}
		\boldsymbol{\sigma}_{ij} = \mathbf{C}_{ijkl} \boldsymbol{\bm{\varepsilon}}_{kl}
		\end{equation}
	where $\boldsymbol{\sigma}_{ij}$ is the 3-D stress tensor, $\boldsymbol{\bm{\varepsilon}}_{kl}$ is the 3-D strain tensor, and $\mathbf{C}_{ijkl}$ is the material specific elastic stif{}fness tensor which is a function of temperature (Table \ref{tab:Ti_stif{}fness}). When the load is released, the material returns to its original state.

\begin{figure}[h]
		\centering
		\includegraphics[width=0.2\textwidth]{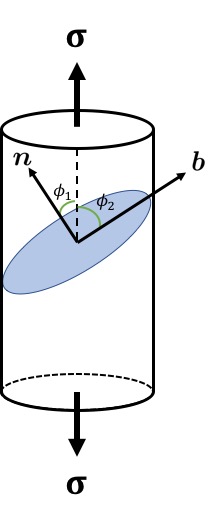}
		\caption{Schematic for the calculation of resolved shear stress for a single crystal. }
		\label{fig:RSS_calc}
	\end{figure}

	The stress tensor can be projected onto each slip system to calculate the resolved shear stress (RSS) using $\tau_{RSS}=\boldsymbol{b \sigma n}$, where $\boldsymbol{b}$ is the unit vector for the slip direction and $\boldsymbol{n}$ is the unit vectors for the slip plane (Fig. \ref{fig:RSS_calc}). Deformation occurs when the RSS reaches the critical resolved shear stress (CRSS), and the material yields.

\begin{table}[h]
	\centering
	\begin{tabular}{c|c|c|c|c|c}
		\textbf{Temp. ($\degree$C)} & $\mathbf{c_{11}}$ & $\mathbf{c_{33}}$  & $\mathbf{c_{44}}$ & $\mathbf{c_{13}}$ & $\mathbf{c_{12}}$ \\ \hline
		\rowcolor{Gray}
		\hline
		25 & 162.4 & 180.7 & 46.7 & 69.0 & 92.0 \\ 
		\rowcolor{White}
		50 & 160.9 &  179.5 & 46.2 & 69.1 & 92.5 \\ 
		\rowcolor{Gray}
		100 &  157.9 & 177.4 & 45.3 & 69.4 & 93.4 \\ 
		\rowcolor{White}
		150 & 155.1 & 175.3 & 44.4 & 69.5 & 94.3 \\ 
		\rowcolor{Gray}
		200 & 152.2 & 173.4 & 43.4 & 69.5 & 95.2 \\ 
		\rowcolor{White}
		250 & 149.5 & 171.5 & 42.4 & 69.2 & 96.1 \\ 
		\rowcolor{Gray}
		300 & 146.8 & 169.6 & 41.4 & 69.2 & 96.7 \\ 
		\rowcolor{White}
		350 & 144.2 & 167.8 & 40.3 & 69.1 & 97.3 \\ 
		\rowcolor{Gray}
		400 & 141.6 & 166.1 & 39.2 & 69.0 & 97.8 \\ 
		\rowcolor{White}
		450 & 139.2 & 164.4 & 38.1 & 69.2 & 98.3 \\ 
		\rowcolor{Gray}
		500 & 136.8 & 162.7 & 37.0 & 68.8 & 98.5 \\ 
		\rowcolor{White}
		550 & 134.5 & 161.0 & 35.9 & 68.8 & 98.8 \\ 
		\rowcolor{Gray}
		600 & 132.2 & 159.3 & 34.8 & 68.8 & 99.1 \\ 
		\rowcolor{White}
		650 & 129.9 & 157.6 & 33.7 & 68.8 & 99.2 \\ 
		\rowcolor{Gray}
		700 & 127.6 & 156.0 & 32.6 &  & 99.3 \\ 
		\rowcolor{White}
		750 & 125.3 & 154.5 & 31.6 &  & 99.4 \\ 
		\rowcolor{Gray}
		800 & 123.1 & 152.9 & 30.7 &  & 99.6 \\ 
		
	\end{tabular}
	\caption{Temperature dependent stif{}fness moduli for Ti (GPa) from Fisher et al. \cite{Fisher1964}.}
	\label{tab:Ti_stif{}fness}
\end{table}

\newpage

\end{appendices}
\newpage
\clearpage

\bibliographystyle{elsarticle-num}
\bibliography{References}

\end{document}